\documentclass[12pt]{article}

\usepackage{graphicx}
\usepackage{amsfonts}
\usepackage{amssymb}
\usepackage{amsthm}
\usepackage{amsmath}
\usepackage{graphicx}
\usepackage{epsfig}
\usepackage{color}

\theoremstyle{plain}
\newtheorem{theorem}{Theorem} 
\newtheorem{lemma}[theorem]{Lemma}
\newtheorem{proposition}[theorem]{Proposition}
\newtheorem{corollary}[theorem]{Corollary}

\newtheorem{claim}[theorem]{Claim}

\theoremstyle{definition}
\newtheorem{definition}{Definition}

\theoremstyle{remark}
\newtheorem{remark}{Remark}
\newtheorem{example}{Example}

\definecolor{red}{named}{Red}
\definecolor{blue}{named}{BlueViolet}
\definecolor{green}{named}{LimeGreen}

\newcommand{\calN}{\mathcal{N}}
\newcommand{\calV}{\mathcal{V}}
\newcommand{\calF}{\mathcal{F}}
\newcommand{\calS}{\mathcal{S}}
\newcommand{\calM}{\mathcal{M}}

\newcommand{\calRs}{\mathcal{R}_{\square}}

\newcommand{\Z}{\mathbb{Z}}
\newcommand{\Reals}{\mathbb{R}}

\newcommand{\defined}{\stackrel{\triangle}{=}}

\newcommand{\hsx}{\ensuremath{ H_*^X }}
\newcommand{\hsy}{\ensuremath{ H_*^Y }}
\newcommand{\sx}{\ensuremath{ \sigma_X^2 }}
\newcommand{\sy}{\ensuremath{ \sigma_Y^2 }}
\newcommand{\sn}{\ensuremath{ \sigma_N^2 }}
\newcommand{\mmse}{\text{mmse}}
\newcommand{\snr}{\text{SNR}}
\newcommand{\I}{\text{I}}
\newcommand{\der}{\text{d}}
\newcommand{\Var}{\text{Var}}
\begin{document}
\title{Scanning and Sequential Decision Making for Multi-Dimensional Data - Part II: the Noisy Case\thanks{The material in this
paper was presented in part at the IEEE International Symposium on Information Theory, Seattle, Washington, United States, July 2006, and accepted to the IEEE International Symposium on Information Theory, Nice, France, June 2007.}}
\author{Asaf Cohen\thanks{Asaf Cohen and Neri Merhav are with the Department of the Electrical Engineering, Technion - I.I.T., Haifa 32000, Israel. E-mails: \{soofsoof@tx,merhav@ee\}.technion.ac.il.}, Tsachy Weissman\thanks{Tsachy Weissman is with the Department of Electrical Engineering, Stanford University, Stanford, CA 94305, USA. E-mail: tsachy@stanford.edu.} and Neri Merhav$^\dagger$}
\maketitle
\date
\abstract{We consider the problem of sequential decision making on random fields corrupted by noise. In this scenario, the decision maker observes a noisy version of the data, yet judged with respect to the clean data. In particular, we first consider the problem of sequentially scanning and filtering noisy random fields. In this case, the sequential filter is given the freedom to choose the path over which it traverses the random field (e.g., noisy image or video sequence), thus it is natural to ask what is the best achievable performance and how sensitive this performance is to the choice of the scan. We formally define the problem of scanning and filtering, derive a bound on the best achievable performance and quantify the excess loss occurring when non-optimal scanners are used, compared to optimal scanning and filtering.

We then discuss the problem of sequential scanning and prediction of noisy random fields. This setting is a natural model for applications such as restoration and coding of noisy images. We formally define the problem of scanning and prediction of a noisy multidimensional array and relate the optimal performance to the clean scandictability defined by Merhav and Weissman. Moreover, bounds on the excess loss due to sub-optimal scans are derived, and a universal prediction algorithm is suggested.

This paper is the second part of a two-part paper. The first paper dealt with sequential decision making on noiseless data arrays, namely, when the decision maker is judged with respect to the same data array it observes.}

\section{Introduction}\label{sec. intro.}Consider the problem of sequentially scanning and filtering (or predicting) a multidimensional noisy data array, while minimizing a given loss function. Particularly, at each time instant $t$, $1 \leq t \leq |B|$, where $|B|$ is the number of sites (``pixels") in the data array, the sequential decision maker chooses a site to be visited, denoted by $\Psi_t$. In the filtering scenario, it first observes the value at that site, and then gives an estimation for the underlying clean value. In the prediction scenario, it is required to give a prediction for that (clean) value, before the actual observation is made. In both cases, both the location $\Psi_t$ and the estimation or prediction may depend on the previously observed values - the values at sites $\Psi_1$ to $\Psi_{t-1}$. The goal is to minimize the cumulative loss after scanning the entire data array. 

Applications of this problem can be found in image and video processing, such as filtering or predictive coding. In these applications, one wishes to either enhance or jointly enhance and code a given image. The motivation behind a prediction/compression-based approach, is that the prediction error may
consist mainly of the noise signal, while the clean signal is recovered by
the predictor. For example, see \cite{Natarajan_Konstantinides_Herley98}. It is clear that different scanning patterns of the image may result in different filtering or prediction errors, thus, it is natural to ask what is the performance of the optimal scanning strategy, and what is the loss when non-optimal strategies are used.

The problem of scanning multidimensional data arrays also arises in other areas of image processing, such as one-dimensional wavelet \cite{Lamarque_Robert96} or median \cite{Krzyzak_et_al01} processing of images, where one seeks a space-filling curve which facilitates the one-dimensional signal processing of the multidimensional data. Other examples include digital halftoning \cite{Velho_Gomes91}, where a space filling curve is sought in order to minimize the effect of false contours, and pattern recognition \cite{Skubalska-Rafajlowicz01}. Yet more applications can be found in multidimensional data query \cite{Asano_et_al97} and indexing \cite{Moon_et_al01}, where multidimensional data is stored on a one-dimensional storage device, hence a locality-preserving space-filling curve is sought in order to minimize the number of continuous read operations required to access a multidimensional object, and rendering of three-dimensional graphics \cite{Bogomjakov_Gotsman02}, \cite{Niedermeier_et_al02}.

An information theoretic discussion of the scanning problem was initiated by Lempel and Ziv in \cite{Lempel_Ziv86}, where the Peano-Hilbert scan was shown to be optimal for compression of individual images. In \cite{Mer_Weiss03}, Merhav and Weissman formally defined a ``scandictor", a scheme for sequentially scanning and predicting a multidimensional data array, as well as the ``scandictability" of a random field, namely, the best achievable performance for scanning and prediction of a random field. Particular cases where this value can be computed and the optimal scanning order can be identified were discussed in that work. One of the main results of \cite{Mer_Weiss03} is the fact that if a stochastic field can be represented autoregressively (under a specific scan $\Psi$)
with a maximum-entropy innovation process, then it is optimally scandicted
in the way it was created (i.e., by the specific scan $\Psi$ and its
corresponding optimal predictor). A more comprehensive survey can be found in \cite{Cohen_Merhav_Weissman_I06} and \cite{Cohen_07}.
In \cite{Cohen_Merhav_Weissman_I06}, the problem of \emph{universal} scanning and prediction of noise-free multidimensional arrays was investigated. Although this problem is fundamentally different from its one-dimensional analogue (for example, one cannot compete successfully with any two scandictors on any individual image), a universal scanning and prediction algorithm which achieves the scandictability of any stationary random field was given, and the excess loss incurred when non-optimal scanning strategies are used was quantified.
 
In \cite{Weiss_Mer_Somekh01}, Weissman, Merhav and Somekh-Baruch, as well
as Weissman and Merhav in \cite{Weiss_Mer01} and \cite{Weiss_Mer04},
extended the problem of universal prediction to the case of a \emph{noisy}
environment. Namely, the predictor observes a noisy version of the
sequence, yet, it is judged with respect to the clean sequence. In this paper, we extend the results of \cite{Mer_Weiss03} and \cite{Cohen_Merhav_Weissman_I06} to this noisy scenario. We formally define the problem of sequentially filtering or predicting a multidimensional data array. First, we derive lower bounds on the best achievable performance. We then discuss the scenario where non-optimal scanning strategies are used. That is, we assume that, due to implementation constraints, for example, one cannot use the optimal scanner for a given data array, and is forced to use an arbitrary scanning order. In such a scenario, it is important to understand what is the excess loss incurred, compared to optimal scanning and filtering (or prediction). We derive upper
bounds on this excess loss. Finally, we briefly mention how the results of \cite{Cohen_Merhav_Weissman_I06} can be exploited in order to construct universal schemes to the noisy case as well. While many of the results for noisy scandiction are extendible from the noiseless case, similarly as results for noisy prediction were extended from results for noiseless prediction \cite{Weiss_Mer01}, the scanning and filtering problem poses new challenges and requires the use of new tools and techniques.

The paper is organized as follows. Section \ref{sec. prob. form.} includes a precise formulation of the problem. Section \ref{sec. filtering} includes the results on scanning and filtering of noisy data arrays, while Section \ref{sec. prediction} is devoted to the prediction scenario. In both sections, particular emphasis is given to the important cases of Gaussian random fields corrupted by Additive White Gaussian Noise (AWGN), under the squared error criterion, and binary random fields corrupted by a Binary Symmetric Channel (BSC), under the Hamming loss criterion.
 
In particular, in Section \ref{sec. Filt. Noisy Data Arrays}, a new tool is used to derive a lower bound on the optimum scanning and filtering performance (Section \ref{sec. Prediction of Noisy Data Arrays} later shows how this tool can be used to strengthen the results of \cite{Mer_Weiss03} in the noise-free scenario as well). Section \ref{subsec. bounds} gives upper bounds on the excess loss in non-optimal scanning. In Section \ref{sec. bounds filtering Gaussian}, the results of Duncan \cite{Duncan70} as well as those of Guo, Shamai and Verd\'u \cite{Guo_Shamai_Verdu05} are used to derive the bounds when the noise is Gaussian, and Section \ref{sec. scantering excess loss binary} deals with the binary setting. Section \ref{sec. universal scantering} uses recent results by Weissman \emph{et}.\ \emph{al}.\ \cite{Weiss_et_al07} to describe how universal scanning and filtering algorithms can be constructed. In the noisy scandiction section, Section \ref{sec. Prediction of Noisy Data Arrays} relates the best achievable performance in this setting, as well as the achieving scandictors, to the \emph{clean scandictability} of the noisy field. Section \ref{sec. universal scandiction noisy} introduces a universal scandiction algorithm, and Section \ref{sec. bounds scandiction} gives an upper bound on the excess loss. In both Section \ref{sec. filtering} and Section \ref{sec. prediction}, the sub-sections describing the optimum performance, the excess loss bounds and the universal algorithms are not directly related and can be read independently. Finally, Section \ref{sec. conc} contains some concluding remarks.

\section{Problem Formulation}\label{sec. prob. form.}
We start with a formal definition of the problem. Let $A$ denote the alphabet, which is either discrete or the real line. Let $N$ be the noisy observation alphabet. Let $\Omega = (A\times N)^{\Z^2}$ be the
observation space (the results can be extended to any finite dimension). A probability measure $Q$ on $\Omega$ is stationary
if it is invariant under translations $\tau_i$, where for each $\omega \in \Omega$ and $i,j \in \Z^2$, $\tau_i(\omega)_j=\omega_{j+i}$ (namely, stationarity means shift invariance). Denote by $\calM(\Omega)$ and $\calM_S(\Omega)$ the
sets of all probability measures on $\Omega$ and stationary probability
measures on $\Omega$, respectively. Elements of $\calM(\Omega)$, \emph{random fields}, will be denoted by upper case letters while elements of $\Omega$, \emph{individual data arrays}, will be denoted by the corresponding lower case. It will also be beneficial to refer to the clean and noisy random fields separately, that is, $\{X_t\}_{t \in \Z^2}$ represents the clean signal and $\{Y_t\}_{t \in \Z^2}$ represents the noisy observations, where for
$t \in \Z^2$, $X_t$ is the random variable corresponding to $X$ at site
$t$.

Let $\calV$ denote the set of all finite subsets of $\Z^2$. For $V \in
\calV$, denote by $X_V$ the restrictions of the data array $X$ to $V$. Let $\calRs$ be the set of all rectangles of the form $V=\Z^2 \cap
([m_1,m_2]\times[n_1,n_2])$. As a special case, denote by $V_n$ the square
$\{0,\ldots,n-1\} \times \{0,\ldots,n-1\}$. For $V \subset \Z^2$, let the interior radius of
$V$ be
\begin{equation}
R(V) \defined \sup\{r: \exists c \text{ s.t. } B(c,r) \subseteq V\},
\end{equation}
where $B(c,r)$ is a closed ball (under the $l_1$-norm) of radius $r$
centered at $c$. Throughout, $\ln(\cdot)$ will denote the natural
logarithm.
\begin{definition}\label{def. scanterer}
A \emph{scanner-filter pair} for a finite set of sites $B \in \calV$ is the following pair $(\Psi,F)$:
\begin{itemize}
\item The scan $\{\Psi_t\}_{t=1}^{|B|}$ is a sequence of
measurable mappings, $\Psi_t: N^{t-1} \mapsto B$ determining the site
to be visited at time $t$, with the property that
\begin{equation}
\Big\{\Psi_1,\Psi_2(y_{\Psi_1}),\Psi_3(y_{\Psi_1},y_{\Psi_2}),\ldots,
\Psi_{|B|}\left(y_{\Psi_1},\ldots,
y_{\Psi_{|B|-1}}\right)\Big\}=B,
\quad \forall y \in N^B.
\end{equation}
\item $\{\tilde{F}_t\}_{t=1}^{|B|}$ is a sequence of
measurable filters, where for each $t$, $\tilde{F}_t: N^{t} \mapsto D$ determines the reconstruction for the value at the site visited at time $t$, based on the current and previous observations, and $D$ is the reconstruction alphabet.
\end{itemize}
\end{definition}
Note that both the scanner $\Psi$ and the filters $\{\tilde{F}_t\}$ base their decisions only on the noisy observations. In the prediction scenario (i.e., noisy scandiction), we define $F_t: N^{t-1} \mapsto D$, that is, $\{F_t\}$ represents measurable predictors, which have access only to previous observations. We allow \emph{randomized} scanner-filter pairs, namely, pairs such that
$\{\Psi_t\}_{t=1}^{|B|}$ or $\{\tilde{F}_t\}_{t=1}^{|B|}$ can be chosen randomly
from some set of possible functions. It is also important to note that we consider only scanners for finite sets of sites, ones which can be viewed merely as a reordering of the sites in a finite set $B$.

The cumulative loss of a scanner-filter pair $(\Psi,\tilde{F})$ up to time $t \leq |B|$ is denoted by
$L_{(\Psi,\tilde{F})}(x_B,y_B)_t$,
\begin{equation} \label{def. L(X,Y)_t}
L_{(\Psi,\tilde{F})}(x_B,y_B)_t =
\sum_{i=1}^{t}{l(x_{\Psi_i},\tilde{F}_i(y_{\Psi_1},\ldots,y_{\Psi_{i}}))},
\end{equation}
where $l:A \times D \mapsto [0,\infty)$ is the loss function. The sum of the instantaneous losses over the
entire data array $B$, $L_{(\Psi,\tilde{F})}(x_B,y_B)_{|B|}$, will be abbreviated as $L_{(\Psi,\tilde{F})}(x_B,y_B)$.

For a given loss function $l$ and a field $Q \in \calM (\Omega)$ restricted
to $B$, define the best achievable scanning and filtering performance by
\begin{equation} \label{def. tildeU(l,Q_B)}
\tilde{U}(l,Q_B) = \inf_{(\Psi,\tilde{F}) \in \mathcal{S}(B)} E_{Q_B} \frac{1}{|B|}
L_{(\Psi,\tilde{F})}(X_B,Y_B),
\end{equation}
where $Q_B$ is the marginal probability measure restricted to $B$ and $\calS(B)$ is the set of \emph{all} possible scanner-filter pairs for
$B$. The best achievable performance for the field $Q$, $\tilde{U}(l,Q)$, is defined by
\begin{equation} \label{def. tildeU(l,Q)}
\tilde{U}(l,Q) = \lim_{n \rightarrow \infty}\tilde{U}(l,Q_{V_n}),
\end{equation}
if this limit exists.

In the prediction scenario, $F_t$ is allowed to base its estimation only on $y_{\Psi_1},\ldots,y_{\Psi_{t-1}}$, and we have
\begin{equation} \label{def. L(X,Y) prediction}
L_{(\Psi,F)}(x_B,y_B) =
\sum_{t=1}^{|B|}{l(x_{\Psi_t},F_t(y_{\Psi_1},\ldots,y_{\Psi_{t-1}}))},
\end{equation}
\begin{equation} \label{def. barU(l,Q_B)}
\bar{U}(l,Q_B) = \inf_{(\Psi,F)} E_{Q_B} \frac{1}{|B|}
L_{(\Psi,F)}(X_B,Y_B),
\end{equation}
and
\begin{equation} \label{def. barU(l,Q)}
\bar{U}(l,Q) = \lim_{n \rightarrow \infty}\bar{U}(l,Q_{V_n}),
\end{equation}
if this limit exists.

The following proposition asserts that for any stationary random field both the limit in \eqref{def. tildeU(l,Q)} and the limit in \eqref{def. barU(l,Q)} exist.
%
\begin{proposition} \label{prop. existance of the limit}
For any stationary field $Q \in \calM_S(\Omega)$ and for any sequence
$\{B_n\}$, $B_n \in \calRs$, satisfying $R(B_n) \rightarrow \infty$, the
limits in \eqref{def. tildeU(l,Q)} and \eqref{def. barU(l,Q)} exist and satisfy
\begin{eqnarray}
\tilde{U}(l,Q)=\lim_{n \rightarrow \infty} \tilde{U}(l,Q_{B_n})=\inf_{\Delta \in \calRs}
\tilde{U}(l,Q_\Delta),
\\
\bar{U}(l,Q)=\lim_{n \rightarrow \infty} \bar{U}(l,Q_{B_n})=\inf_{\Delta \in \calRs}
\bar{U}(l,Q_\Delta).
\end{eqnarray}
\end{proposition}
Since $\tilde{U}(l,Q_B)$ and $\bar{U}(l,Q_B)$, possess the sub-additivity property, e.g., for any $V, V', V \cap V' = \emptyset$, there exists a scanner-filter pair $(\Psi,\tilde{F})$ (or a scandictor $(\Psi,F)$) on $V \cup V'$ such that
\begin{equation}
E_Q L_{(\Psi,\tilde{F})}(X_{V \cup V'},Y_{V \cup V'}) \leq |V|\tilde{U}(l,Q_{V}) +
|V'|\tilde{U}(l,Q_{V'}),
\end{equation}
the proof of Proposition \ref{prop. existance of the limit} follows
verbatim that of \cite[Theorem 1]{Mer_Weiss03}.

\section{Filtering of Noisy Data Arrays}\label{sec. filtering}
In this section, we consider the scenario of scanning and filtering. In this case, a lower bound on the best achievable performance is derived. For the cases of Gaussian random fields corrupted by AWGN and binary valued fields observed through a BSC, we derive bounds on the excess loss when a non-optimal scanner is used (with an optimal filter). Finally, we briefly discuss universal scanning and filtering. 
\subsection{A Lower Bound on the Best Achievable Scanning and Filtering Performance}\label{sec. Filt. Noisy
  Data Arrays}
We assume an invertible memoryless channel, meaning the channel input
distribution of a single symbol is uniquely determined given the output
distribution. As an example, a discrete memoryless channel with an invertible channel matrix can be kept in mind. See \cite{Weissman_et_al05} for a discussion on the conditions on the channel matrix for the invertibility property to hold. Moreover, as will be elaborated on later, the result below applies to more general channels, including continuous ones. 

In the case of an invertible channel, we define associated Bayes envelope by 
\begin{equation}\label{def. flp}
f_l(P)=\min_{g(\cdot)}El(X,g(Y)),
\end{equation}
where $P$ is the distribution of the channel output $Y$. Define
\begin{equation}\label{def. zd}
\zeta(d)=\max\{H(P):f_l(P) \leq d\},
\end{equation}
and let $\bar{\zeta}(\cdot)$\label{def. bzd} be the upper concave ($\cap$) envelope of
$\zeta(\cdot)$. 
\begin{theorem}\label{theo. scantering performance}
Let $Y_B$ be the output of an invertible memoryless channel whose input
is $X_B$. Then, for any scanner-filter pair $(\Psi,\tilde{F})$ we have
\begin{equation}
\bar{\zeta}\left(\frac{1}{|B|}E_{Q_B}L_{(\Psi,\tilde{F})}(X_B,Y_B)\right) \geq \frac{1}{|B|}H(Y_B),
\end{equation}
that is,
\begin{equation}\label{eq. in. theorem scantering per.}
\bar{\zeta}\left(\tilde{U}(l,Q_B)\right) \geq \frac{1}{|B|}H(Y_B).
\end{equation}
\end{theorem}
\begin{proof}{}
We prove the above theorem for the discrete case. Yet, the derivations below apply to the continuous case as well, with summations replaced by the appropriate integrals and the entropy replaced by differential entropy. 

Denote by $\Psi(Y_B)$ the reordered output sequence, that is,
$\{Y_{\Psi_1},Y_{\Psi_2},\ldots,Y_{\Psi_{|B|}}\}$. We have,
\begin{eqnarray}
H(Y_B) & \stackrel{(a)}{=} & H(\Psi(Y_B))
\nonumber\\
&=& \sum_{t=1}^{|B|}{H(Y_{\Psi_t}|Y^{\Psi_{t-1}})}
\nonumber\\
&=& \sum_{t=1}^{|B|}{\sum_{y^{\Psi_{t-1}}}H(Y_{\Psi_t}|Y^{\Psi_{t-1}}=y^{\Psi_{t-1}})P(y^{\Psi_{t-1}})}
\nonumber\\
& \stackrel{(b)}{\leq} & \sum_{t=1}^{|B|}{\sum_{y^{\Psi_{t-1}}}\zeta\left(E_{Q_B}\left\{l\left(X_{\Psi_t},\tilde{F}_{t}(y^{\Psi_{t-1}},Y_{\Psi_t})\right)|Y^{\Psi_{t-1}}=y^{\Psi_{t-1}}\right\}\right)P(y^{\Psi_{t-1}})}
\nonumber\\
& \stackrel{(c)}{\leq} &  \sum_{t=1}^{|B|}{\sum_{y^{\Psi_{t-1}}}\bar{\zeta}\left(E_{Q_B}\left\{l\left(X_{\Psi_t},\tilde{F}_{t}(y^{\Psi_{t-1}},Y_{\Psi_t})\right)|Y^{\Psi_{t-1}}=y^{\Psi_{t-1}}\right\}\right)P(y^{\Psi_{t-1}})}
\nonumber\\
& \stackrel{(d)}{\leq} & \sum_{t=1}^{|B|}{\bar{\zeta}\left(E_{Q_B}l\left(X_{\Psi_t},\tilde{F}_t(Y^{\Psi_t})\right)\right)}
\nonumber\\
& \stackrel{(e)}{\leq} & |B|\bar{\zeta}\left(\frac{1}{|B|}E_{Q_B}L_{\tilde{F}}\left(\Psi(X_B),\Psi(Y_B)\right)\right)
\nonumber\\
&=& |B|\bar{\zeta}\left(\frac{1}{|B|}E_{Q_B}L_{(\Psi,\tilde{F})}(X_B,Y_B)\right).\label{eq. proof of scantering bound}
\end{eqnarray}
The equality $(a)$ is since the reordering does not change the entropy of $Y_B$. While this is clear for data-independent reordering, more caution is required when $\Psi$ is a data-dependent scan. Yet, this can be proved using the chain rule, and noting that conditioned on $Y_{\Psi_1}^{\Psi_{t-1}}$, the next site $\Psi_t$ is fixed (this is similar to the proof of \cite[Proposition 13]{Cohen_Merhav_Weissman_I06}). The inequalities (b) and (c)
follow from the definitions of $\zeta$ and $\bar{\zeta}$ respectively,
and (d) and (e) follow from Jensen's inequality.
\end{proof}
At this point, a few remarks are in order. Theorem \ref{theo. scantering performance} is the direct analogue of the lower bounds in \cite{Mer_Weiss03} for the filtering scenario. Note, however, that it holds for any finite set of sites $B$. Furthermore, it applies to arbitrarily distributed random fields (even non-stationary fields), and to a wide family of loss functions. In fact, the only condition on $l(\cdot,\cdot)$ is that the associated Bayes envelope $f_l(P)$ is well defined. Note also that the lower bound on $\tilde{U}(l,Q)$ given in Theorem \ref{theo. scantering performance} results from the application of a \emph{single letter} function, $\bar{\zeta}^{-1}(\cdot)$, to the normalized entropy of the noisy field, $\frac{1}{|B|}H(Y_B)$. That is, the memory in $(X_B,Y_B)$ is reflected only in $\frac{1}{|B|}H(Y_B)$.

The proof of Theorem \ref{theo. scantering performance} is general and direct, however, it lacks the insightful geometrical interpretation which led to the lower bound in \cite{Mer_Weiss03}. Therein, Merhav and Weissman showed that the transformation from a data array to an error sequence (defined by a specific scandictor $(\Psi,F)$) is volume preserving. Thus, the least expected cumulative error is the \emph{radius of a sphere}, whose volume is the volume of the set of all typical data arrays of the source. This happens when all the typical data arrays of the source map to a sphere in the ``error vectors" space, and thus Merhav and Weissman were able to identify cases where the lower bound is tight. Currently, we cannot point out specific cases in which \eqref{eq. in. theorem scantering per.} is tight. Moreover, as the next two examples show, in the scanning and filtering scenario (unlike the scanning and prediction scenario we discuss in Section \ref{sec. prediction}), $\zeta(d)$ may not be concave, and thus $\zeta(d) \ne \bar{\zeta}(d)$. Note, in this context, that there is no natural \emph{time sharing} solution in this case, as there is no natural trade-off between two (or more) optimal points, and there is only one criterion to be minimized - the cumulative scanning and filtering loss (as opposed to rate versus distortion, for example). 
\subsubsection{Binary Input and BSC}
To illustrate its use, we specialize Theorem \ref{theo. scantering performance} to the case of binary input through a BSC, i.e., the input random field $X_{V_n}$ is binary, and $Y_{V_n}$ is the output of a BSC whose input is $X_{V_n}$ and crossover probability is $\delta < 1/2$. Note, however, that although the derivations below are specific for binary alphabet and Hamming loss, they are easily extendible to arbitrary finite alphabet and discrete memoryless channel with a channel transition matrix $\Pi$ and loss function $\Lambda(\cdot,\cdot)$.

\begin{figure}
\centering
\includegraphics[scale=0.6]{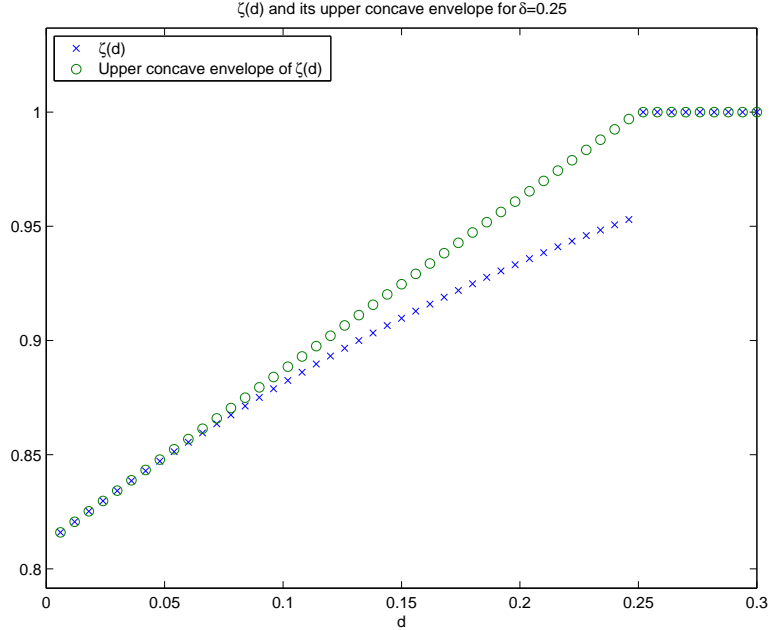}
\caption{The function $\zeta(d)$, as it appears in \eqref{eq. def. zeta binary}, and its upper concave envelope, $\bar{\zeta(d)}$, both plotted for $\delta=0.25$. Note that $\zeta(d)$ and $\bar{\zeta}(d)$ have analytic expressions, and the plots are discrete only to better distinguish between them.}
\label{fig_zeta}
\end{figure}

To compute the lower bound on the best achievable scanning and filtering performance, we evaluate $f_l(P)$ and $\zeta(d)$. By the definitions in \eqref{def. flp} and \eqref{def. zd}, we consider the scalar problem of estimation of a random variable $X$ based on its noisy observation $Y$. Denote by $p_Y$ the probability $P(Y=1)$ and by $p_X$ the probability $P(X=1)$. The best achievable performance, $f_{l_H}(p_Y)$, which clearly depends on $\delta$, and, hence, denoted $f_{\delta}(p_Y)$, is given by
\begin{eqnarray}
f_{\delta}(p_Y) &=& \sum_{x,y}{P(x,y)l_H\left(x,g_{opt}(y)\right)}
\nonumber\\
&=& \sum_{y}{P(y)\sum_{x}{P(x|y)l_H\left(x,g_{opt}(y)\right)}}
\nonumber\\
&\stackrel{(a)}{=}& \sum_y{P(y)\min_x P(x|y)}
\nonumber\\
&=& \sum_y{\min_x P(x,y)}
\nonumber\\
&=& \min\left\{p_X(1-\delta), \delta(1-p_X) \right\} + \min\left\{p_X \delta, (1-\delta)(1-p_X) \right\}
\nonumber\\
&=& \min\{p_X,1-p_X,\delta\}
\nonumber\\
&\stackrel{(b)}{=}& \min\left\{\frac{p_Y-\delta}{1-2\delta},\frac{1-p_Y-\delta}{1-2\delta},\delta\right\},\label{eq. computing f_d}
\end{eqnarray}
where (a) results from the optimality of $g_{opt}(y)$ and (b) results from the invertability of the channel. Consequently,
\begin{eqnarray}\label{eq. def. zeta binary}
\zeta(d) &=& \max_p h_b(p) \quad \text{s.t.} \quad f_{\delta}(p) \leq d
\nonumber\\
&=& \Bigg\{ 
\begin{array}{ll}
h_b(\delta * d) & d < \delta \\
1 & d \ge \delta,
\end{array} 
\end{eqnarray}  
where $h_b(\cdot)$ is the binary entropy function and $\delta * d = d(1-\delta)+\delta(1-d)$. Note that since $\delta * \delta < 1/2$ for $0 < \delta < 1/2$, there is a discontinuity at $d=\delta$, hence $\zeta(d)$ is generally not concave and $\bar{\zeta}(d) \ne \zeta(d)$ (although $\bar{\zeta}(d)$ can be easily calculated). Figure \ref{fig_zeta} includes plots of both $\zeta(d)$ and $\bar{\zeta(d)}$ for $\delta=0.25$. We also mention that $d=\delta$ is a realistic cumulative loss in non-trivial situations, as there are cases where ``say-what-you-see" (and thus suffer a loss $\delta$) is the best any filter can do \cite{Orden_Weiss06}. Furthermore, note that $\zeta(d)$ is not the maximum entropy function $\gamma(d)$ used in \cite{Mer_Weiss03} to derive the lower bound on the scandictability.

Finally, exact evaluation of the bound in  Theorem \ref{theo. scantering performance} may be difficult in many cases, as the entropy $\frac{1}{|B|}H(Y_B)$ may be hard to calculate, and only bounds on its value can be used.\footnote{Think, for example, of an input process which is a first order Markov source. While the entropy rate of the input is known, the output is a hidden Markov process whose entropy rate is unknown in general.} At the end of Section \ref{sec. scantering excess loss binary}, we give a numerical example for the bound in Theorem \ref{theo. scantering performance} using a lower bound on the entropy rate.
\begin{remark}
Clearly, $\zeta(d)$ is interesting only in the region $d \leq \delta$, as any reasonable filter will have an expected normalized cumulative loss smaller or equal to the channel crossover probability. However, due to the discontinuity at $d=\delta$, $\zeta(d)$ is concave for $d < \delta$ but not for $d \leq \delta$. This is fortunate, as if $\zeta(d)$ was concave on $d \leq \delta$, Theorem \ref{theo. scantering performance} would have resulted in $h_b(\delta*\delta)$ as an upper bound on the entropy rate of any binary source corrupted by a BSC, which is erroneous (for example, it violates $h_b(\pi * \delta)$ as a lower bound on the entropy rate of a first order Markov source with transition probability $\pi$ corrupted by a BSC with crossover probability $\delta$).
\end{remark}
\subsubsection{Gaussian Channel}
Consider now the case where $Y_{V_n}$ is the output of an AWGN channel, whose input is arbitrarily distributed. Assume the squared error loss. As the optimal filter is clearly the conditional expectation, $\zeta(d)$ in this case is given by
\begin{equation}
\zeta(d)=\max \left\{H(X+N): \Var(X|X+N) \leq d\right\},
\quad N \sim \calN(0,\sigma_N^2), \quad N \perp X.
\end{equation}
Since $H(X+N|X)=H(N)$ is fixed, this is similar to the classical Gaussian channel capacity problem, only now the input constraint is $\Var(X|X+N) \leq d$, which generally depends on the \emph{distribution} of $X$ rather than solely on its variance, and hence is not necessarily achieved by Gaussian $X$.

When the input is also limited to be Gaussian, however, the optimization problem in \eqref{def. zd} is trivial and $\zeta(d)$ can be easily calculated (note that in this case $\zeta(d)$ is valid only to bound the performance for scanning and filtering of Gaussian fields corrupted by AWGN). Since the distributions depend only on the variance (assuming zero expectation), we have $f_{l_s}(P) = f_{l_s}(\sy)$, and, in fact,
\begin{eqnarray}\label{fls}
f_{l_s}(\sy) &=& \frac{\sn \sx}{\sn + \sx}
\nonumber\\
&=& \sn - \frac{\sigma_N^4}{\sy}.
\end{eqnarray} 
Hence,
\begin{eqnarray}
\zeta(d) &=& \max \frac{1}{2}\ln(2 \pi e \sy) \quad \text{s.t.} \quad f_{l_s}(\sy) \leq d
\nonumber\\
&=&\Bigg\{ 
\begin{array}{ll}
 \frac{1}{2}\ln\left(2 \pi e \frac{\sigma_N^4}{\sn-d}\right)& d < \sn \\
\infty & d \ge \sn.
\end{array}
\end{eqnarray} 
Unlike the binary setting, here the cumulative loss, $d$, will be strictly smaller than $\sn$ for any non-trivial setting and reasonable filter, as the error in symbol-by-symbol filtering is $\sn \frac{\sx}{\sx+\sn} < \sn$. Yet, $\zeta(d)$ is convex ($\cup$) for $d < \sn$, and the chain of inequalities in \eqref{eq. proof of scantering bound} cannot be tight. 
\subsection{Bounds on the Excess Loss of Non-Optimal Scanners}\label{subsec. bounds}
Theorem \ref{theo. scantering performance} gives a lower bound on the optimum scanning and filtering performance. However, it is interesting to investigate what is the excess scanning and filtering loss when non-optimal scanners are used. Specifically, in this section we address the following question: Suppose that, for practical reasons for example, one uses a non-optimal scanner, accompanied with the optimal filter for that scan. How large is the excess loss incurred by this scheme with respect to optimal scanning and filtering?

We consider both the case of a Gaussian channel and squared error loss (with Gaussian or arbitrarily distributed input) and the case of a binary source passed through a BSC and Hamming loss. While the tools we use in order to construct such a bound for the binary case are similar to the ones used in \cite{Cohen_Merhav_Weissman_I06}, we develop a new set of tools and techniques for the Gaussian setting.
\subsubsection{Gaussian Channel}\label{sec. bounds filtering Gaussian}
We investigate the excess scanning and filtering loss when non-optimal scanners are used, for the case of arbitrarily distributed input corrupted by a Gaussian channel. We first focus attention on the case where the input is Gaussian as well, and then derive a new results for the more general setting. 
 
Similarly as in \cite{Cohen_Merhav_Weissman_I06}, the bound is achieved by bounding the absolute difference between the scanning and filtering performance of any two scans, $\Psi^1$ and $\Psi^2$, assuming both use their optimal filters. This bound, however, results from a relation between the performance of \emph{discrete time} filtering and \emph{continuous time} filtering, together with the fundamental result of Duncan \cite{Duncan70} on the relation between mutual information and causal minimal mean square error estimation in a Gaussian channel. Namely, we use the \emph{mutual information in continuous time} as a scan invariant feature, and the actual value of the excess loss bound results from the difference between discrete and continuous time filtering problems, as will be made precise below. 

From now on we assume the loss function is the squared error loss, $l_s(\cdot)$.
 We start with several definitions. Let $X$ be a Gaussian random variable, $X \sim \calN(0,\sx)$. Consider the following two estimation problems:
\begin{itemize}
\item The scalar problem of estimating $X$ based on $Y=X+N$, where $N \sim \calN(0,\sn)$, independent of $X$. 
\item The continuous time problem of causally estimating $X_t \equiv X$, $t \in [0,1]$, based on $Y_t$, which is an AWGN-corrupted version of $X_t$, the Gaussian noise having a spectral density level of $\sn$.
\end{itemize}
To bound the sensitivity of the scanning and filtering performance, it is beneficial to consider the difference between the estimation errors in the above two problems, that is,
\begin{equation}\label{eq. what we want to bound}
\int_0^1 \Var(X_t|Y^t) \der t - \Var(X|Y),
\end{equation}
where $Y^t$ is the continuous time signal $\{Y_{t'}\}_{t'=0}^t$. Clearly, $\Var(X|Y) = \frac{\sx \sn}{\sx + \sn}$. Since $\int_{0}^t{Y_{t'}}\der t'$ is a sufficient statistics in the estimation of $X_t \equiv X$, $\Var(X_t|Y^t)$ is equivalent to the squared error in estimating $X$ based on $X+\tilde{N}$, $\tilde{N}$ being a Gaussian random variable, independent of $X$, with zero mean and variance $\sn/t$. Thus,
\begin{eqnarray}\label{def. f}
\int_0^1 \Var(X_t|Y^t) \der t - \Var(X|Y) &=& \int_0^1 \frac{\sx (\sn / t)}{\sx + (\sn / t) } \der t- \frac{\sx \sn}{\sx + \sn} 
\nonumber\\
&=& \sn \ln \left(1+\frac{\sx}{\sn}\right) - \frac{\sx \sn}{\sx + \sn}
\nonumber\\
&=& \sn f\left(\frac{\sx}{\sn}\right),
\end{eqnarray}
where
\begin{equation}\label{def. f scantering}
f(x)=\ln(1+x)-\frac{x}{x+1}.
\end{equation}
The following is the main result in this sub-section.
\begin{theorem}\label{theo. bound on sens. of scantering Gaussian}
Let $X_{V_n}$ be a Gaussian random field with a constant marginal distribution satisfying $\Var(X_i)=\sx < \infty$ for all $i \in V_n$. Let $Y_i=X_i+N_i$, where $N_{V_n}$ is a white Gaussian noise of variance $\sigma_N^2$, independent of $X_{V_n}$. Then, for any two scans $\Psi^1$ and $\Psi^2$, we have
\begin{equation}\label{eq. first bound}
\frac{1}{n^2}\left|EL_{(\Psi^1,\tilde{F}^{opt})}\left(X_{V_n},Y_{V_n}\right)-EL_{(\Psi^2,\tilde{F}^{opt})}\left(X_{V_n},Y_{V_n}\right)\right| 
\leq \sn f\left(\frac{\sx}{\sn}\right). 
\end{equation}
\end{theorem}
Theorem \ref{theo. bound on sens. of scantering Gaussian} bounds the absolute difference between the scanning and filtering performance of any two scanners, $\Psi^1$ and $\Psi^2$, assuming they use their optimal filters. Clearly, since the scanners are arbitrary, this result can also be interpreted as the difference in performance between any scan $\Psi$, and the best achievable performance, $\tilde{U}(l,Q_{V_n})$. Note that the bound value, $\sn f\left(\frac{\sx}{\sn}\right)$, is a single letter expression, which depends on the input field $X_{V_n}$ and the noise $N_{V_n}$ only through their variances. Namely, the bound does not depend on the \emph{memory in $X_{V_n}$}.
\begin{proof}[Proof (Theorem \ref{theo. bound on sens. of scantering Gaussian})]
As mentioned earlier, the comparison between any two scans is made by bounding the normalized cumulative loss of \emph{any scan} $\Psi$ in terms of a \emph{scan invariant} entity, which is the mutual information. 

For simplicity, assume first that the scan $\Psi$ is data-independent, namely, it is merely a reordering of the entries of $Y_{V_n}$. In this case, $\{X_{\Psi_i}\}_{i=1}^{n^2}$ is a discrete time Gaussian vector. We construct from it a \emph{continuous time process}, $\{X_t^{(c)}\}_{t \in [0,n^2]}$, where for any $t \in [i-1,i)$, $X_{t}^{(c)}=X_{\Psi_i}$, $i \in \{1,2,\ldots,n^2\}$. That is, $X_t^{(c)}$ is a piecewise constant process, whose constant values at intervals of length $1$ correspond to the original values of the discrete time vector $\{X_{\Psi_i}\}$. Let $\{Y_{\Psi_i}\}$ and $\{Y_t^{(c)}\}$ be the AWGN-corrupted versions on $\{X_{\Psi_i}\}$ and $X_t^{(c)}$, namely, $Y_{\Psi_i}=X_{\Psi_i}+N_{\Psi_i}$ and $Y_t^{(c)}$ is constructed according to
\begin{equation}
\der Y_t^{(c)} = X_t^{(c)} \der t + \sigma_N \der  W_t,\quad t \in [0,n^2],
\end{equation}
where $W_t$ is a standard Brownian motion. Observe that the white Gaussian noise, $\sigma_N \der  W_t$, has a spectral density of level $\sn$, similar to the variance of the discrete time noise $N_{V_n}$. Since we switch from discrete time to continuous time, it is important to note that the noise value in the two problems is equivalent. That is, if the discrete time field $X_{V_n}$ is corrupted by noise of variance $\sn$, then we wish the continuous time white noise to have a spectrum such that the integral over an interval of
length $1$, whose integrand is the continuous output $Y_t^{(c)}$ (and thus is a
sufficient statistics in order to estimate the piecewise-continuous
input $X_t$ in this interval), will be a random variable which is exactly
$X_{\Psi_i} + N_{\Psi_i}$, $N_{\Psi_i}$ having a variance of $\sn$.
 
We have,
\begin{align} 
\frac{1}{n^2}E_{Q_{V_n}}&L_{(\Psi,\tilde{F}^{opt})}\left(X_{V_n},Y_{V_n}\right)
\nonumber\\ 
&= \frac{1}{n^2}\sum_{i=1}^{n^2}{\Var\left(X_{\Psi_i}|Y_{\Psi_1}^{\Psi_{i}}\right)}
\nonumber\\
&\stackrel{(a)}{=} \frac{1}{n^2}\sum_{i=1}^{n^2}\Bigg[\int_0^1{\Var\left(X_{\Psi_i}|Y_{\Psi_1}^{\Psi_{i-1}},\{Y_{t'}^{(c)}\}_{{t'} \in [i-1,i-1+t]}\right)\der t}  
\nonumber\\
& \qquad -\sn f\left(\frac{\Var\big(X_{\Psi_i}|Y_{\Psi_1}^{\Psi_{i-1}}\big)}{\sn}\right)\Bigg]
\nonumber\\
&\stackrel{(b)}{\ge} \frac{1}{n^2}\sum_{i=1}^{n^2}\int_0^1{\Var\left(X_{\Psi_i}|Y_{\Psi_1}^{\Psi_{i-1}},\{Y_{t'}^{(c)}\}_{t' \in [i-1,i-1+t]}\right)\der t}  
 -\sn f\left(\frac{\Var\big(X_1\big)}{\sn}\right)
\nonumber\\
& \stackrel{(c)}{=}\frac{1}{n^2}2\sn I \left(\{X_t^{(c)}\}_{t \in [0,n^2]};\{Y_t^{(c)}\}_{t \in [0,n^2]}\right)  
 -\sn f\left(\frac{\Var\big(X_1\big)}{\sn}\right)
\nonumber\\
& =\frac{1}{n^2}2\sn I \left(\{X_{\Psi_i}\};\{Y_{\Psi_i}\}\right) - \sn f\left(\frac{\Var\big(X_1\big)}{\sn}\right)
\nonumber\\
& \stackrel{(d)}{=} \frac{1}{n^2}2\sn I \left(X_{V_n};Y_{V_n}\right) - \sn f\left(\frac{\Var\big(X_1\big)}{\sn}\right).\label{eq. proof of scantering bound G}
\end{align}
The equality (a) is from the application of \eqref{def. f} with $X=X_{\Psi_i}|Y_{\Psi_1}^{\Psi_{i-1}}$, i.e., with $X_{\Psi_i}$ distributed conditioned on $Y_{\Psi_1}^{\Psi_{i-1}}$. Note conditioned on $Y_{\Psi_1}^{\Psi_{i-1}}$, $X_{\Psi_i}$ is indeed Gaussian, and that \eqref{def. f} applies to \emph{any} Gaussian $X$ corrupted by Gaussian noise. The inequality (b) is since $\Var(X_{\Psi_i}|Y_{\Psi_1}^{\Psi_{i-1}}) \leq \Var(X_1)$ and due to the increasing monotonicity of $f$, (c) is since the resulting integral from $0$ to $n^2$ is simply the minimal mean square error in filtering $\{Y_t^{(c)}\}$ (as $Y_{\Psi_i}$ is a sufficient statistics with respect to $\{Y_{t'}^{(c)}\}_{t' \in [i-1,i-1+t]}$), and the application of Duncan's result \cite[Theorem 3]{Duncan70}. Finally, (d) is since the mutual information is invariant to the reordering of the random variables. To complete the proof of Theorem \ref{theo. bound on sens. of scantering Gaussian}, simply note that since $f(x)$ is non-negative for $x>0$, by (a) above, the normalized cumulative loss can be upper bounded as well, that is,
\begin{equation}
\frac{1}{n^2}E_{Q_{V_n}}L_{(\Psi,\tilde{F}^{opt})}\left(X_{V_n},Y_{V_n}\right) \leq \frac{1}{n^2}\sum_{i=1}^{n^2}\int_0^1{\Var\left(X_{\Psi_i}|Y_{\Psi_1}^{\Psi_{i-1}},\{Y_{t'}^{(c)}\}_{t' \in [i-1,i-1+t]}\right)\der t}
\end{equation}
hence, similarly as in the chain of inequalities leading to \eqref{eq. proof of scantering bound G},
\begin{equation}\label{eq. upper bound like GSV}
\frac{1}{n^2}E_{Q_{V_n}}L_{(\Psi,\tilde{F}_{opt})}\left(X_{V_n},Y_{V_n}\right) \leq \frac{1}{n^2}2\sn I \left(X_{V_n};Y_{V_n}\right).
\end{equation} 
In fact, equation \eqref{eq. upper bound like GSV} can be viewed as the scanning and filtering analogue of \cite[eq. (156a)]{Guo_Shamai_Verdu05}.

Now, if the scan $\Psi$ is data-dependent, the above derivations apply, with the use of the smoothing property of conditional expectation. That is, conditioned on $Y_{\Psi_1}^{\Psi_{i-1}}$, the position $\Psi_i$ is fixed (assuming deterministic scanners, though random scanning order can be tackled with a similar method), relation (a) in \eqref{eq. proof of scantering bound G} holds since it holds conditioned on $Y_{\Psi_1}^{\Psi_{i-1}}$, and relation (c) holds as the mutual information is invariant under data-dependent reordering as well. This is very similar to the methods used in the proof of \cite[Proposition 13]{Cohen_Merhav_Weissman_I06}, where it was shown that the entropy of a vector is invariant to data-dependent reordering.
\end{proof}

At this point, a few remarks are in order. A very simple bound, applicable to arbitrarily distributed fields and under squared error loss (yet interesting mainly in the Gaussian regime) results from noting that for \emph{any} random variables $X$ and $Y=X+N$,
\begin{equation}\label{eq. bound for arbit. X for scantering sens.}
0 \leq \Var(X|Y) \leq \sn \frac{\sx}{\sx+\sn}.
\end{equation}
Namely, simple symbol by symbol restoration results in a cumulative loss of at most $\sn \frac{\sx}{\sx+\sn}$, and we have, 
\begin{eqnarray}
\frac{1}{n}EL_{(\Psi,\tilde{F}^{opt})}\left(\{X_i\}_{i=1}^{n},\{Y_i\}_{i=1}^{n}\right)
& =& \frac{1}{n}\sum_i \Var(X_{\Psi_i}|Y_{\Psi_1}^{\Psi_{i}})
\nonumber\\
& \leq &\frac{1}{n}\sum_i \Var(X_{\Psi_i}|Y_{\Psi_{i}})
\nonumber\\
&=& \sn \frac{\sx}{\sx+\sn}.\label{eq. upper bound on scantering performance with var}
\end{eqnarray}
Thus, the excess loss in non-optimal scanning cannot be greater than that value, hence,
\begin{equation}\label{eq. first bound in cor}
\frac{1}{n^2}\left|EL_{(\Psi^1,\tilde{F}^{opt})}\left(X_{V_n},Y_{V_n}\right)-EL_{(\Psi^2,\tilde{F}^{opt})}\left(X_{V_n},Y_{V_n}\right)\right| 
\leq \sn \frac{\sx}{\sx+\sn}. 
\end{equation}
\begin{figure}
\centering
\includegraphics[scale=0.6]{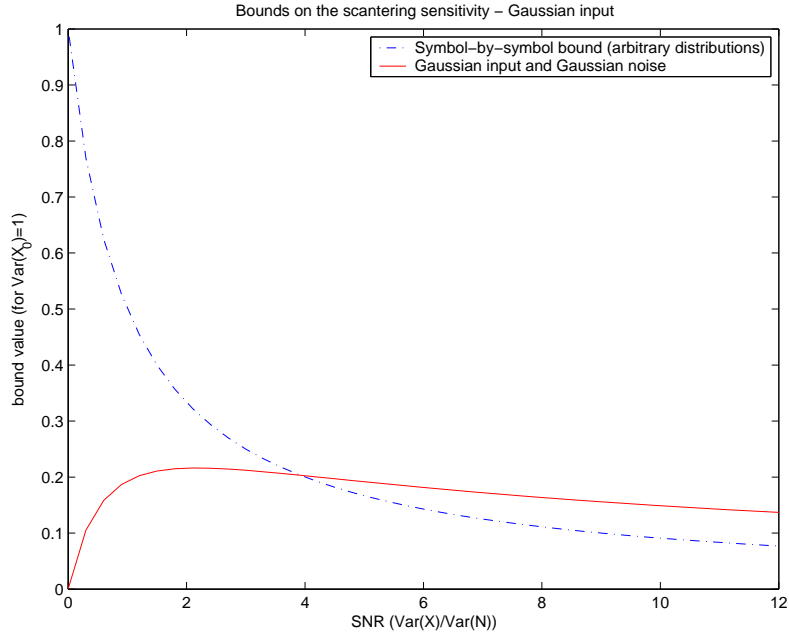}
\caption{Bounds on the excess loss in scanning and filtering of Gaussian input corrupted by AWGN. The solid line is the bound given in Theorem \ref{theo. bound on sens. of scantering Gaussian}. The dashed line is the bound given in \eqref{eq. first bound in cor}.}
\label{fig_sim}
\end{figure}
In the next sub-section, we derive a tighter bound than the bound in \eqref{eq. first bound in cor}, applicable to arbitrarily distributed noise-free fields. However, since this bound may be harder to evaluate, it is interesting to discuss the properties of \eqref{eq. first bound in cor} as well.

Both the bound in Theorem \ref{theo. bound on sens. of scantering Gaussian} and the bound in \eqref{eq. first bound in cor} are in the form of $\Var(X_1)g(\snr)$, for some $g$, where $\snr = \sx / \sn$\label{snr}. This means that any bound obtained for a certain SNR applies to all values of $\Var(X_1)$ by rescaling. The bound in Theorem \ref{theo. bound on sens. of scantering Gaussian} has the form $\Var(X_1)\frac{f(\snr)}{\snr}$, where $f(\cdot)$ was defined in \eqref{def. f scantering}, and we have
\begin{equation}
\lim_{\snr \to 0^+} \frac{f(\snr)}{\snr}=\lim_{\snr \to \infty} \frac{f(\snr)}{\snr}=0,
\end{equation}
that is, the scan is inconsequential at very high or very low SNR. This is clear as at high SNR the current observation is by far the most influential, and whatever previous observations used is inconsequential. For low SNR, the cumulative loss is high whatever the scan is. Unlike the bound in Theorem \ref{theo. bound on sens. of scantering Gaussian}, \eqref{eq. first bound in cor} does not predict the correct behavior for $\snr \to 0^+$, and is mainly interesting in the high SNR regime. 

The above observations are also evident in Figure \ref{fig_sim}, which includes both the bound given in Theorem \ref{theo. bound on sens. of scantering Gaussian}, applicable to Gaussian fields, and \eqref{eq. first bound in cor}, applicable to arbitrarily distributed fields. It is also evident that in the case of Gaussian fields, $\frac{f(\snr)}{\snr}$ has a unique maximum of approximately $0.216$, that is, the excess loss due to a suboptimal scan at any SNR is upper bounded by $0.216 \Var(X_1)$. 
\begin{remark}
It is clear from the proof of Theorem \ref{theo. bound on sens. of scantering Gaussian} that an upper bound on the expression in \eqref{eq. what we want to bound}, valid for arbitrarily distributed input $X$, may yield an upper bound on the excess scanning and filtering loss which is also valid for arbitrarily distributed random fields. However, while the integral in \eqref{eq. what we want to bound} can be upper bounded by assuming a Gaussian $X$, $\Var(X|Y)$ has no non-trivial lower bound. In fact, in \cite{Guo04}, it is shown that if $X$ is the following binary random variable,
\begin{equation}
X = \left\{ \begin{array}{ll}
\sqrt{\frac{1-p}{p}} & \textrm{w.p. }p,\\
-\sqrt{\frac{p}{1-p}} & \textrm{w.p. }1-p,\\
\end{array} \right.
\end{equation}
for which $EX=0$ and $EX^2=1$, then we have
\begin{equation}
Var(X|Y) \leq \frac{1}{2p(1-p)}e^{-\frac{\sigma_X^2/\sigma_N^2}{4p(1-p)}},
\end{equation}
which can be arbitrarily close to $0$ for small enough $p$. Thus, the only lower bound on $V(X|Y)$ which is valid for any $X$ with $EX^2 < \infty$, and depends only on $\sigma_X^2$ and $\sigma_N^2$, is $0$ (and hence results is a bound weaker than Theorem \ref{theo. bound on sens. of scantering Gaussian} or \eqref{eq. first bound in cor}).
\end{remark}

In the next two subsections, we derive new bounds on the excess loss, which are valid for more general input fields. First, we generalize the bound in Theorem \ref{theo. bound on sens. of scantering Gaussian}. While the result may be complex to evaluate in its general form, we show that for binary input fields the bound admits a simple form. We then show that if the input alphabet is continuous, then a non-trivial bound on $\Var(X|Y)$ can be derived easily, which, in turn, results in a new bound on the excess loss.  

%
\noindent\textbf{A Generalization of Theorem \ref{theo. bound on sens. of scantering Gaussian}.}
A generalization of Theorem \ref{theo. bound on sens. of scantering Gaussian} results from revisiting equality (a) of \eqref{eq. proof of scantering bound G}, which is simply the application of \eqref{def. f} with $X=X_{\Psi_i}|Y_{\Psi_1}^{\Psi_{i-1}}$. While it is clear that an expression similar to that in \eqref{def. f} can be computed for non-Gaussian $X$, it is not clear that $X_{\Psi_i}|Y_{\Psi_1}^{\Psi_{i-1}}$ has the same distribution for any $1 \leq i \leq n^2$ (unlike the Gaussian setting, where $X_{\Psi_i}|Y_{\Psi_1}^{\Psi_{i-1}}$ is always Gaussian). Nevertheless, using the definition below, one can generalize Theorem \ref{theo. bound on sens. of scantering Gaussian} for arbitrarily distributed inputs as follows.

For any $(X_{V_n},Y_{V_n})$, where $Y_{V_n}$ is the AWGN-corrupted version of $X_{V_n}$, define
\begin{equation}\label{def. f*}
f^*(X_{V_n},\sn) = \max_{\Psi,1 \leq i \leq n^2}\left\{ \int_0^1 \Var\left(X_{\Psi_i}|Y_{\Psi_1}^{\Psi_{i-1}},\{Y_{t'}^{(c)}\}_{t' \in [i-1,i-1+t]}\right) \der t - \Var\left(X_{\Psi_i}|Y_{\Psi_1}^{\Psi_{i}}\right) \right\}.
\end{equation}
%
\begin{theorem}\label{theo. bound on sens. of scantering Arbitrary}
Let $X_{V_n}$ be an arbitrarily distributed random field, with a constant marginal distribution satisfying $\Var(X_i)=\sx < \infty$ for all $i \in V_n$. Let $Y_i=X_i+N_i$, where $N_{V_n}$ is a white Gaussian noise of variance $\sigma_N^2$, independent of $X_{V_n}$. Then, for any two scans $\Psi^1$ and $\Psi^2$, we have
\begin{equation}\label{eq. generalized bound}
\frac{1}{n^2}\left|EL_{(\Psi^1,\tilde{F}^{opt})}\left(X_{V_n},Y_{V_n}\right)-EL_{(\Psi^2,\tilde{F}^{opt})}\left(X_{V_n},Y_{V_n}\right)\right| 
\leq f^*(X_{V_n},\sn). 
\end{equation}
\end{theorem}
The proof of Theorem \ref{theo. bound on sens. of scantering Arbitrary} is similar to that of Theorem \ref{theo. bound on sens. of scantering Gaussian}, and appears in Appendix \ref{app. proof of bound on sens. of scantering Arbitrary}. 

Note that $f^*(X_{V_n},\sn)$ is scan-independent, as it includes a maximization over all possible scans. At first sight, it seems like this maximization may take the sting out of the excess loss bound. However, as the example below shows, at least for the interesting scenario of binary input, this is not the case.

\begin{figure}
\centering
\includegraphics[scale=0.6]{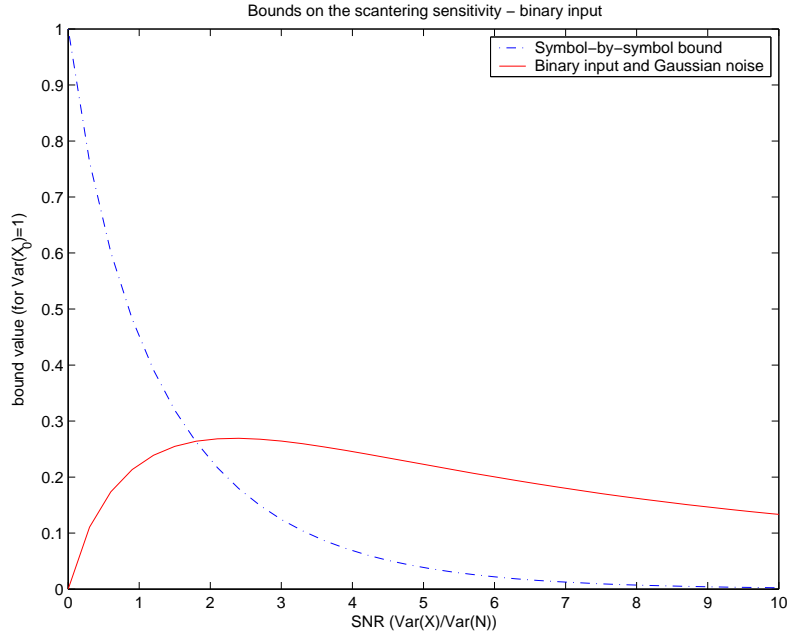}
\caption{Bounds on the excess loss in scanning and filtering of binary input fields corrupted by AWGN. The solid line is the bound given in \eqref{eq. sens bound binary input Gaussian noise} (that is, Theorem \ref{theo. bound on sens. of scantering Arbitrary}), and the dashed line is the symbol-by-symbol bound given in \eqref{eq. sens bound binary input Gaussian noise simple}.}
\label{fig_sens_binary_gaussian}
\end{figure}

First, however, a few more general remarks are in order. Since important insight can be gained when using the results of Guo, Shamai and Verd\'u \cite{Guo_Shamai_Verdu05}, let us mention the setting used therein. In \cite{Guo_Shamai_Verdu05}, one wishes to estimate $X$ based on $\sqrt{\snr}X+N$, where $N$ is a standard normal random variable. Denote by $\I(\snr)$ and $\mmse(\snr)$ the mutual information between $X$ and $\sqrt{\snr}X+N$, and the minimal mean square error in estimating $X$ based on $\sqrt{\snr}X+N$, respectively. Note that $\Var(X|Y)$ in our setting equals $\sx \mmse(\sx/\sn)$. Under these definitions,
\begin{equation}
\frac{\der}{\der \snr}\I(\snr) = \frac{1}{2}\mmse(\snr),
\end{equation}
or, equivalently, 
\begin{equation}
\I(\snr)=\frac{1}{2}\int_{0}^{\snr}{\mmse(\gamma)}\der \gamma.
\end{equation}
Consequently, the result of Theorem \ref{theo. bound on sens. of scantering Gaussian} can be restated as 
\begin{equation}\label{eq. the I-mmse form}
\frac{1}{n^2}\left|EL_{(\Psi^1,\tilde{F}^{opt})}\left(X_{V_n},Y_{V_n}\right)-EL_{(\Psi^2,\tilde{F}^{opt})}\left(X_{V_n},Y_{V_n}\right)\right| 
\leq 2\sn\I(\snr)-\sx\mmse(\snr), 
\end{equation}
where $\I(\snr)=\frac{1}{2}\ln(1+\snr)$ and $\mmse(\snr)=\frac{1}{1+\snr}$ are simply the mutual information and minimal mean square error of the \emph{scalar} problem (hence, a single letter expression) of estimating a Gaussian $X$ based on $\sqrt{\snr}X+N$, where $N$ standard Gaussian. In fact, the bound in Theorem \ref{theo. bound on sens. of scantering Arbitrary} will always have the form $2\sn\I(\snr)-\sx\mmse(\snr)$, for some $X^*$ whose distribution is the maximizing distribution in \eqref{def. f*}. The next example shows that this is indeed the case for binary input as well, and the resulting bound can be easily computed. 
\begin{example}[\textit{Binary input and AWGN}]
Consider the case where $X_{V_n}$ is a binary random field, with a symmetric marginal distribution (that is, $P(X_0=\sigma_X)=P(X_0=-\sigma_X)=1/2$). Note that the $X_i$'s are not necessarily i.i.d., and any dependence between them is possible. $Y_{V_n}$ is the AWGN-corrupted version of $X_{V_n}$. To evaluate the bound in Theorem \ref{theo. bound on sens. of scantering Arbitrary}, $f^*(X_{V_n},\sn)$ should be calculated. However, for any scan $\Psi$ and time $i$, $X_{\Psi_i}|Y_{\Psi_1}^{\Psi_{i-1}}$ is still a binary random variable, taking the values $\pm \sigma_X$ with probabilities $(p,1-p)$, for some $0 \leq p \leq 1/2$. Hence,
\begin{equation}\label{eq. f^* for binary}
f^*(X_{V_n},\sn) \leq \max_{0 \leq p \leq 1/2}\left\{\int_0^1 \Var(X_t|Y^t) \der t - \Var(X|Y)\right\},
\end{equation}
where $X$ is a binary random variable, taking the values $\pm \sigma_X$ with probabilities $(p,1-p)$, $X_t \equiv X$, $Y=X+N$ and $Y_t$ is the AWGN-corrupted version of $X_t$. The following result holds for any random variable $X$. 
\begin{claim}\label{claim monotonicity}
For any random variable $X$ with $\Var(X)=\sx < \infty$, the expression in \eqref{eq. what we want to bound} is monotonically increasing in $\sx$.
\end{claim}
\begin{proof}
We have,
\begin{align}
\int_0^1 \Var(X|Y^t) \der t - \Var(X|Y) &= \int_{0}^{1}{\sx \mmse(\sx/\sn t)\der t}-\sx \mmse(\sx/\sn)
\nonumber\\
&=\sn \int_{0}^{\sx/\sn}{\mmse(\gamma)\der \gamma}-\sx \mmse(\sx/\sn).
\end{align}
Thus,
\begin{align}
\frac{\der}{\der\sx}\left(\int_0^1 \Var(X_t|Y^t) \der t - \Var(X|Y)\right) &= -\sx \frac{\der}{\der\sx}\mmse(\sx/\sn)
\nonumber\\
&= -2 \snr \frac{\der^2}{\der\snr^2}\I(\snr)
\nonumber\\
& \geq 0,
\end{align}
where the last inequality is by \cite[Corollary 1]{Guo_Shamai_Verdu05}.
\end{proof}
Claim \ref{claim monotonicity} simply states that the monotonicity of $f(\cdot)$ used in inequality (b) of \eqref{eq. proof of scantering bound G} is not specific for Gaussian input, and holds for any $X$. Thus, by Claim \ref{claim monotonicity}, the term in the braces of equation \eqref{eq. f^* for binary} is monotonically increasing in the variance of $X$, which is simply $4 \sx (p-p^2)$. Thus, it is maximized by $p=1/2$, and we have
\begin{equation}\label{eq. sens bound binary input Gaussian noise}
f^*(X_{V_n},\sn) = 2\sn\I(\snr)-\sx\mmse(\snr),
\end{equation}
where $\I(\snr)$ and $\mmse(\snr)$ are the mutual information and minimal mean square error in the estimation of $X$ based on $\sqrt{\snr}X+N$, where $X$ is binary symmetric and $N$ is a standard normal. Since the conditional mean estimate in this problem is $\tanh(\sqrt{\snr}Y)$, we have \cite{Guo_Shamai_Verdu05}
\begin{equation}
\I(\snr)=\snr - \frac{1}{\sqrt{2 \pi}}\int_{-\infty}^{\infty}{e^{-\frac{y^2}{2}}\ln\cosh(\snr-\sqrt{\snr}y)dy},
\end{equation}
and
\begin{equation}\label{def. mmse binary}
\mmse(\snr)=1- \frac{1}{\sqrt{2 \pi}}\int_{-\infty}^{\infty}{e^{-\frac{y^2}{2}}\tanh(\snr-\sqrt{\snr}y)dy},
\end{equation}
so the bound can be computed numerically. The above bound is plotted in Figure \ref{fig_sens_binary_gaussian}. Similarly to the case of Gaussian input, it is insightful to compare this bound to a simple symbol-by-symbol filtering bound. That is, since for any binary $X$ corrupted by AWGN of variance $\sn$, $0 \leq \Var(X|Y) \leq \sx\mmse(\snr)$, we have
\begin{equation}\label{eq. sens bound binary input Gaussian noise simple}
\frac{1}{n^2}\left|EL_{(\Psi^1,\tilde{F}^{opt})}\left(X_{V_n},Y_{V_n}\right)-EL_{(\Psi^2,\tilde{F}^{opt})}\left(X_{V_n},Y_{V_n}\right)\right| 
\leq \sx\mmse(\snr), 
\end{equation}
where $\mmse(\snr)$ is given in \eqref{def. mmse binary}. This is simply the analogue of \eqref{eq. first bound in cor} to the binary input setting.
\end{example}
%

%
\noindent\textbf{A Bound for Arbitrarily Distributed Continuous Input.}
In this sub-section, we derive an additional bound on the excess scanning and filtering loss under squared error. We assume, however, that the input random field $X_{V_n}$ is over $\Reals^{V_n}$, and that $X_i | Y_{V_n}$ has a finite differential entropy for any $i \in V_n$ (roughly speaking, this means that in the denoising problem of $X_i$, $X_i|Y_{V_n}$ is a non-degenerated continuous random variable). Under the above assumptions, we derive an excess loss bound which is not only valid for non-Gaussian input, but also depends on the memory in the random field $(X_{V_n},Y_{V_n})$. On the other hand, it is important to note that the bound below is mainly asymptotic, and may be much harder to evaluate compared to the bounds in Theorem \ref{theo. bound on sens. of scantering Gaussian} or \eqref{eq. first bound in cor}.

By \cite[Theorem 9.6.5]{Cov_Thom91}, for any $X,Y$ with a finite conditional differential entropy $H(X|Y)$, 
\begin{equation}\label{eq. bound on var using entropy}
\Var(X|Y) \geq \frac{1}{2 \pi e} \exp\left\{2H(X|Y)\right\}.
\end{equation} 
Thus,
\begin{eqnarray}
\frac{1}{n}EL_{(\Psi,\tilde{F}^{opt})}\left(X_{V_n},Y_{V_n}\right)
& = & \frac{1}{n^2}\sum_{i=1}^{n^2} \Var(X_{\Psi_i}|Y_{\Psi_1}^{\Psi_{i}})
\nonumber\\
& \stackrel{(a)}{\ge} & \frac{1}{n^2}\sum_{i=1}^{n^2}\frac{1}{2\pi e}\exp\{2H(X_{\Psi_i}|Y_{\Psi_1}^{\Psi_{i}})\}
\nonumber\\
&\stackrel{(b)}{\ge} & \frac{1}{n^2}\sum_{i=1}^{n^2}\frac{1}{2\pi e}\exp\{2H(X_{\Psi_i}|Y_{\Psi_1}^{\Psi_{n^2}})\}
\nonumber\\
&\stackrel{(c)}{\ge} & \frac{1}{2\pi e}\exp\left\{2\frac{1}{n^2}\sum_{i=1}^{n^2} H(X_{\Psi_i}|Y_{\Psi_1}^{\Psi_{n^2}})\right\},\label{eq. lower bound on scantering with diff entropy}
\end{eqnarray}
where (a) is by applying \eqref{eq. bound on var using entropy} with $Y = Y_{\Psi_1}^{\Psi_i}$, (b) is since conditioning reduces entropy and (c) is by applying Jensen's inequality.

The expression $\frac{1}{n^2}\sum_{i=1}^{n^2} H(X_{\Psi_i}|Y_{\Psi_1}^{\Psi_{n^2}})$ equals $\frac{1}{n^2}\sum_{i=1}^{n^2} H(X_{\Psi'_1}|Y_{\Psi'_1}^{\Psi'_{n^2}})$ for any two scanners $\Psi$ and $\Psi'$, since equality holds even without the expectation implicit in the entropy function. Thus, it is scan-invariant.  
Define
\begin{equation}\label{def. H+}
H^+(X|Y) = \liminf_{n \to \infty}\frac{1}{|V_n|}\sum_{i \in V_n} H(X_i|Y_{V_n}).
\end{equation} 
$H^+(X|Y)$ can be seen as the asymptotic normalized entropy in the denoising problem of $\{X\}$ based on its noisy observations. Note that the entropies in \eqref{def. H+} are differential. The following proposition gives a new lower bound on the excess scanning and filtering loss under squared error.
\begin{proposition}\label{prop. excess scantering loss with H+}
Let $X_{V_n}$ be an arbitrarily distributed continuous valued random field with $Var(X_i)=\sx$ for all $i$. Let $Y_i=X_i+N_i$, where $N_{V_n}$ is a white noise of variance $\sigma_N^2$, independent of $X_{V_n}$. Assume that $X_i | Y_{V_n}$ has a finite differential entropy for any $i \in V_n$. Then, for any two scans $\Psi^1$ and $\Psi^2$, we have
\begin{multline}
\liminf_{n \to \infty}\frac{1}{|V_n|}\left|EL_{(\Psi^1,\tilde{F}^{opt})}\left(X_{V_n},Y_{V_n}\right)-EL_{(\Psi^2,\tilde{F}^{opt})}\left(X_{V_n},Y_{V_n}\right)\right| 
\\
\leq \frac{\sn \sx}{\sx+\sn}-\frac{1}{2\pi e}\exp\left\{2H^+(X|Y)\right\}
.
\end{multline}
\end{proposition}
\begin{proof}{}
The proof follows directly by applying the lower bound on the scanning and filtering performance given in \eqref{eq. lower bound on scantering with diff entropy} and the upper bound in \eqref{eq. upper bound on scantering performance with var}.
\end{proof}
The bound in Proposition \ref{prop. excess scantering loss with H+} is always at least as tight as the bound in \eqref{eq. first bound in cor} (and thus tighter than the bound in Theorem \ref{theo. bound on sens. of scantering Gaussian} for high SNR). For example, if 
the estimation error of $X_i$ given $Y_{V_n}$ tends to zero as $n$ increases (as in the case where $X_i = X$ for all $i$), then $\exp\left\{2H^+(X|Y)\right\} =0$. However, if $X_i$ cannot be reconstructed completely by $Y_{V_n}$, then the bound may be tighter than \eqref{eq. first bound in cor}. It is far from being a tight bound on the excess loss, though. In the extreme case where all $X_i$'s are i.i.d., the excess loss bound in Proposition \ref{prop. excess scantering loss with H+} is $\frac{\sn \sx}{\sx+\sn}-\Var(X_1|Y_1) > 0$ (for non Gaussian $X$), while it is clear that all reasonable scanner-filter pairs perform the same. Finally, note that any lower bound on $H^+(X|Y)$ results in an upper bound on the scanning and filtering excess loss. For example, since
\begin{equation}
H^+(X|Y) \ge H(X_0|Y_{-k}^{k},X_{-k-1},X_{k+1})
\end{equation}
for any finite $k$, one can compute a simple upper bound on the excess loss, at least for first order Markov $\{X\}$.
%
\subsubsection{Binary Input and BSC}\label{sec. scantering excess loss binary}
\begin{figure}
\centering
\includegraphics[scale=0.6]{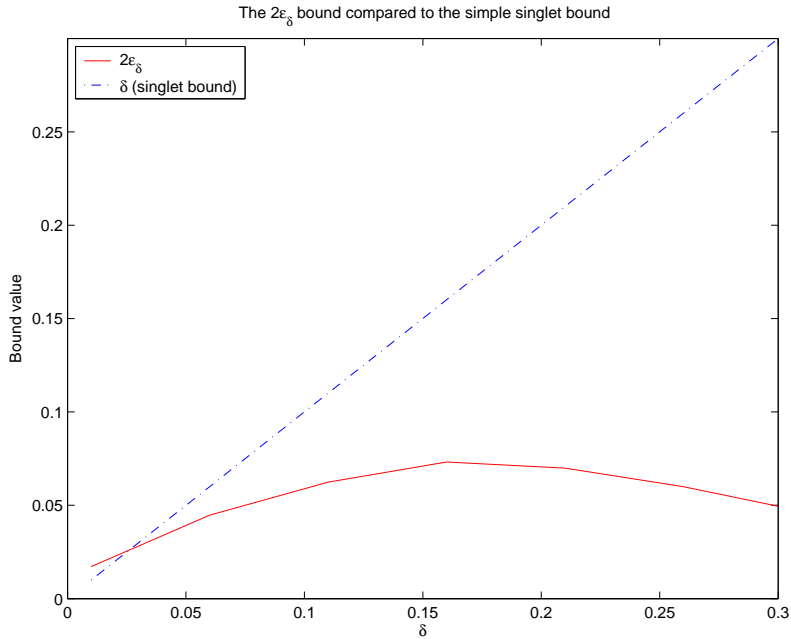}
\caption{Bounds on the excess loss in scanning and filtering of binary random fields corrupted by a BSC. The solid line is the bound in Theorem \ref{theo. non sensitivity to scan in filtering} ($2 \epsilon_\delta$), and the dashed line is the singlet bound ($f_\delta(1/2)=\delta$).}
\label{fig_binary_scantering_bounds}
\end{figure}

Unlike the Gaussian setting discussed in Section \ref{sec. bounds filtering Gaussian}, where the bound on the excess loss resulted from a continuous-time equality, with the \emph{mutual information} serving as the scan-invariant feature, in the case of binary input and a BSC the \emph{entropy} of the random field will play the key role, similar to \cite{Cohen_Merhav_Weissman_I06}. As given in Section \ref{sec. Filt. Noisy
  Data Arrays}, the best achievable performance (in the scalar problem) is given by
\begin{equation}
f_\delta(p)=\min\left\{\frac{p-\delta}{1-2\delta},\frac{1-p-\delta}{1-2\delta},\delta\right\},
\end{equation}
where $p$ is the probability that the channel output is $1$ and $\delta$ is the channel crossover probability. Note that $f_\delta(p)$ is not the Bayes envelope associated with estimating $X_t$ using $Y_t$ under Hamming loss. However, as is clear from the derivations in \eqref{eq. computing f_d}, and will be evident from the proof of the following theorem, $f_\delta(P(y_t|y^{t-1}))$ is the \emph{expectation} of the Bayes envelope (associated with estimating $X_t$ using $Y^t$ under Hamming loss) with respect to the distribution $P(y_t|y^{t-1})$. Define
\begin{equation}\label{def. epsilon_delta}
\epsilon_\delta = \min_{a,b} \max_{\delta \leq p \leq 1/2} \left| ah_b(p)+b - f_\delta(p)\right|.
\end{equation}
The following is the main result in this sub-section. 
\begin{theorem}\label{theo. non sensitivity to scan in filtering}
Let $Y_B$ be the output of a BSC with crossover probability $\delta$ whose input
is $X_B$. Then, for any scanner-filter pair $(\Psi,\tilde{F}^{opt})$, where $\tilde{F}^{opt}$ is the optimal filter for the scan $\Psi$, we have
\begin{equation}\label{eq. loss in non optimal scantering - multidimensional}
\left| \frac{1}{|B|}E_{Q_B}L_{\Psi,\tilde{F}^{opt}}(X_B,Y_B) - \tilde{U}(l_H,Q_B)\right| \leq 2\epsilon_\delta.
\end{equation}
\end{theorem}
Even without evaluating $\epsilon_{\delta}$ explicitly, it is easy to see that the excess loss when using non optimal scanners is quite small in this binary filtering scenario. For example, for $\delta = 0.1$ and $\delta =0.25$ we have $\epsilon_{\delta} < 0.035$ and $\epsilon_{\delta} < 0.03$ respectively, yielding a maximal loss of $0.07$ or even $0.06$. Figure \ref{fig_binary_scantering_bounds} includes the value of $2\epsilon_\delta$ as a function of $\delta$. Similarly to Section \ref{sec. bounds filtering Gaussian}, it is compared to a simple bound on the excess loss which results from simply bounding the Hamming loss of any filter by $0$ from below and $\delta$ from above (namely, $\delta$ is the resulting bound on the excess loss). The values in Figure \ref{fig_binary_scantering_bounds} should also be compared to $0.16$, which is the bound on the excess loss in the clean \emph{prediction} scenario \cite{Cohen_Merhav_Weissman_I06}, or even to larger values in the noisy prediction scenario, to be discussed in the next section. The fact that the filtering problem is less sensitive to the scanning order is quite clear as the noisy observation of $X_{\Psi_t}$ is available under any scan. Finally, it is not hard to show that in the limits of $\delta \to 0$ and $\delta \to 1/2$ (high and low SNR, respectively), we have $\epsilon_\delta \to 0$, which is expected, as the scanning is inconsequential in these cases (note, however, that the singlet bound, $\delta$, does not predict the correct behavior at low SNR).
\begin{proof}[Proof (Theorem \ref{theo. non sensitivity to scan in filtering})]
We first show that for any arbitrarily distributed binary $n$-tuple $X^n$ and any $0 \leq \delta < 1/2$
\begin{equation}\label{eq. loss in non optimal scantering - sequence}
\left| a^*_\delta \frac{1}{n}H(Y^n)+b^*_\delta - \frac{1}{n}EL_{l_H}^{opt}(X^n,Y^n)\right| \leq \epsilon_\delta,
\end{equation}
where $EL_{l_H}^{opt}(X^n,Y^n)$ is the expected cumulative Hamming loss in optimally filtering $X^n$ based on $Y^n$ and $a^*_\delta$ and $b^*_\delta$ are the minimizers of $\epsilon_\delta$ in \eqref{def. epsilon_delta}. Indeed,
\begin{eqnarray}
&&\hspace{-1cm} \left| a^*_\delta \frac{1}{n}H(Y^n)+b^*_\delta - \frac{1}{n}EL_{l_H}^{opt}(X^n,Y^n)\right|
\nonumber\\
&&\hspace{-0.5cm}= \bigg|\frac{1}{n}\sum_{t=1}^{n}\sum_{y^{t-1}}P(y^{t-1})\sum_{x_t,y_t}\bigg[-a^*_\delta P(x_t,y_t|y^{t-1})\ln P(y_t|y^{t-1}) + b^*_\delta P(x_t,y_t|y^{t-1})
\nonumber\\
&&\hspace{+6.5cm} - P(x_t,y_t|y^{t-1}) l_H\left(x_t,\tilde{F}^{opt}(y^t)\right)\bigg]\bigg|
\nonumber\\
&&\hspace{-0.5cm}\leq \frac{1}{n}\sum_{t=1}^{n}\sum_{y^{t-1}}P(y^{t-1}) \bigg| a^*_\delta h_b\left(P(y_t|y^{t-1})\right)+b^*_\delta - \sum_{x_t,y_t} P(x_t,y_t|y^{t-1}) l_H\left(x_t,\tilde{F}^{opt}(y^t)\right) \bigg|
\nonumber\\
&&\hspace{-0.5cm}\leq \frac{1}{n}\sum_{t=1}^{n}\sum_{y^{t-1}}P(y^{t-1}) \bigg| a^*_\delta h_b\left(P(y_t|y^{t-1})\right)+b^*_\delta - \sum_{y_t} P(y_t|y^{t-1})\sum_{x_t}P(x_t|y^t) l_H\left(x_t,\tilde{F}^{opt}(y^t)\right) \bigg|.
\nonumber\\
\end{eqnarray}
Consider the summation $\sum_{y_t} P(y_t|y^{t-1})\sum_{x_t}P(x_t|y^t) l_H\left(x_t,\tilde{F}^{opt}(y^t)\right)$. As $\tilde{F}$ is optimal, the inner sum equals at most $\min\{P(x_t=0|y^t),P(x_t=1|y^t)\}$. Thus, similar to the derivations in \eqref{eq. computing f_d}, we have
\begin{align}
&\sum_{y_t} P(y_t|y^{t-1})\sum_{x_t}P(x_t|y^t) l_H\left(x_t,\tilde{F}^{opt}(y^t)\right)
\nonumber\\
&\hspace{+1cm}=\sum_{y_t} P(y_t|y^{t-1})\min\{P(x_t=0|y^t),P(x_t=1|y^t)\}
\nonumber\\
&\hspace{+1cm}=\min\left\{\frac{P(y_t=0|y^{t-1})-\delta}{1-2\delta},\frac{P(y_t=1|y^{t-1})-\delta}{1-2\delta},\delta\right\}.
\end{align}
Let $p=P(y_t=0|y^{t-1})$. Note that $\min\{p,1-p\} \geq \delta$. We have
\begin{align}
&\left| a^*_\delta \frac{1}{n}H(Y^n)+b^*_\delta - \frac{1}{n}EL_{l_H}^{opt}(X^n,\delta)\right|
\nonumber\\
&\hspace{+1cm} \leq \frac{1}{n}\sum_{t=1}^{n}\sum_{y^{t-1}}P(y^{t-1}) \left| a^*_\delta h_b(p)+b^*_\delta - \min\left\{\frac{p-\delta}{1-2\delta},\frac{1-p-\delta}{1-2\delta},\delta\right\} \right|
\nonumber\\
&\hspace{+1cm} \leq \frac{1}{n}\sum_{t=1}^{n}\sum_{y^{t-1}}P(y^{t-1}) \max_{\delta \leq p \leq 1/2}\left| a^*_\delta h_b(p)+b^*_\delta - f_\delta(p) \right|
\nonumber\\
&\hspace{+1cm} = \epsilon_\delta,
\end{align}
which establishes \eqref{eq. loss in non optimal scantering - sequence}. However, the same inequality can be proved for any reordering of the data $\Psi$ (similar to the proof of \cite[Proposition 13]{Cohen_Merhav_Weissman_I06}), consequently,
\begin{equation}\label{eq. loss in non optimal scantering - YB}
\left| a^*_\delta \frac{1}{|B|}H(\Psi(Y_B))+b^*_\delta - \frac{1}{|B|}E_{Q_B}L_{\Psi,\tilde{F}^{opt}}(X_B,Y_B)\right| \leq \epsilon_\delta.
\end{equation}
Using \eqref{eq. loss in non optimal scantering - YB}, remembering that $H(Y_B)=H(\Psi(Y_B))$ for any $\Psi$, and applying the triangle inequality results in \eqref{eq. loss in non optimal scantering - multidimensional}.
\end{proof}
Note that analogous ideas were used by Verd\'u and Weissman to bound the absolute difference between the denoisability and erasure entropy \cite{Verdu_Weissman_07}.

\begin{figure}
\centering
\includegraphics[scale=0.6]{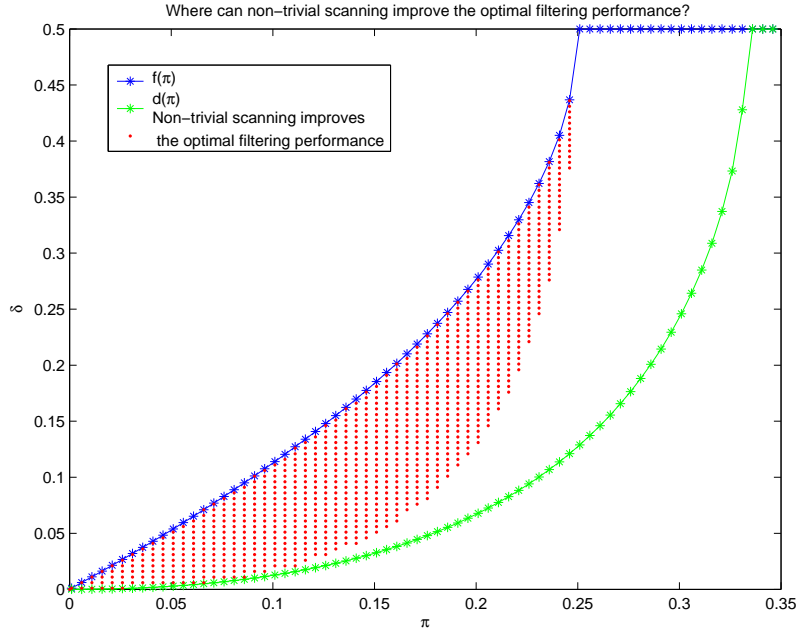}
\caption{Where can a simple (suboptimal) ``odds-then-evens'' scan improve on the trivial scanning order and optimal filtering scheme. $\pi$ is the transition probability of the symmetric, first order, Markov source and $\delta$ is the channel crossover probability.}
\label{fig. hmm}
\end{figure}

\begin{figure}
\centering
\includegraphics[scale=0.8]{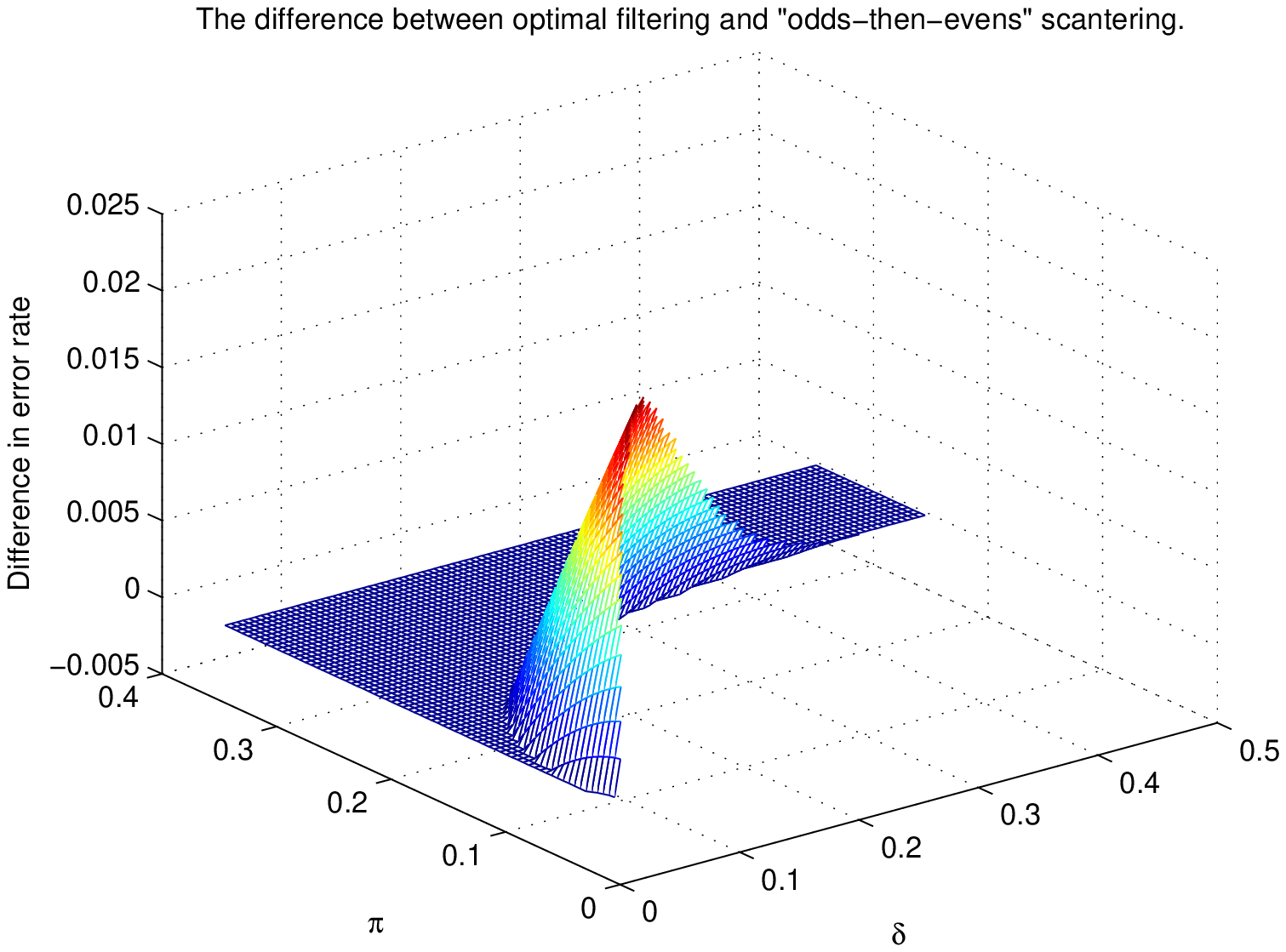}
\caption{The actual difference between the optimal filtering error rate and the ``odds-then-evens" scanning and filtering error rate. $\pi$ is the transition probability of the symmetric, first order, Markov source and $\delta$ is the channel crossover probability. Only values for which $\delta < f(\pi)$ are shown.}
\label{fig. hmm_diff}
\end{figure}

Theorem \ref{theo. scantering performance} gives a lower bound on the best achievable scanning and filtering performance. Theorems \ref{theo. bound on sens. of scantering Gaussian}, \ref{theo. bound on sens. of scantering Arbitrary} and \ref{theo. non sensitivity to scan in filtering} give an upper bound on the maximal possible difference between the normalized cumulative loss of any two scanners (accompanied by the optimal filters), or any one scanner compared to the optimal scan. Although Theorem \ref{theo. scantering performance} is similar to the
results of \cite{Mer_Weiss03}, even for
the relatively simple examples of a Gaussian field through a Gaussian
memoryless channel or a binary source through a binary symmetric
channel we have no results which can parallel \cite[Theorem
17]{Mer_Weiss03} or \cite[Corollary 21]{Mer_Weiss03}, i.e., give an
example of an optimal scanner-filter pair for a certain scenario. However, as the next example shows, we can identify situations when scanning and filtering improves the filtering results, i.e., non trivial scanning of the data results in strictly better restoration. Moreover, the example below illustrates the use of the results derived in this section.  
\begin{example}[\textit{One Dimensional Binary Markov Source and the BSC}] 
In this case, it is not too hard to construct a scheme in which non-trivial scanning improves the filtering performance. In \cite{Orden_Weiss06}, Ordentlich and Weissman study the
optimality of symbol by symbol (singlet) filtering and decoding. That
is, the regions (depending on the source and channel parameters) where a memoryless scheme to
estimate $X_i$ is optimal with respect to causal (filtering) or non
causal (denoising) non-memoryless schemes. Clearly, in the regions where singlet denoising is optimal (a fortiori singlet filtering), scanning cannot improve the filtering performance. However, consider the
region where singlet filtering is optimal, yet singlet denoising is
not. In this region, there exists $k$ for which the estimation
error in estimating $X_i$ based on $Y_{i-k}^{i+k}$ is strictly smaller
than that based on $Y_i$ (as the optimal filter is memoryless yet the optimal denoiser is not). Hence, a scanner which in the first pass scans $k$ contiguous
symbols, then skips one, etc., and in the second pass returns to fill in the
holes, accompanied by singlet filtering in the first pass and
non-memoryless in the second, has strictly better filtering
performance than the trivial scanner.

For a binary symmetric Markov source with a transition probability
$\pi \leq \frac{1}{2}$, corrupted by a BSC with
crossover probability $\delta$, \cite[Corollary 3]{Orden_Weiss06}
asserts that singlet filtering (``say-what-you-see'' scheme in this
case) is optimal if and only if
\begin{equation}
\delta \leq f(\pi) \defined \frac{1}{2}(1-\sqrt{\max\{1-4\pi,0\}}).
\end{equation}
Singlet denoising, on the other hand, is optimal if and only if
\begin{equation}
\delta \leq d(\pi) \defined
\frac{1}{2}\left(1-\sqrt{\max\{1-4\left(\frac{\pi}{1-\pi}\right)^2,0\}}\right).
\end{equation}
Consider a scanner-filter pair which scans the data using an ``odds-then-evens'' scheme. On the odds, ``say-what-you-see''
filtering is used. On the
evens, $Y_{i-1}^{i+1}$ are used in order to estimate $X_i$.\footnote{This is to
have few simple steps in the
forward-backward algorithm \cite[Section 5]{Ephraim_Merhav02} which is required to compute
$P(x_i|y_{i-1}^{i+1})$. The generalization to $Y_{i-k}^{i+k}$ is straightforward.} The results are in Figures
\ref{fig. hmm} and \ref{fig. hmm_diff}. In Figure \ref{fig. hmm}, the points marked with ``x'' are where the
``odds-then-evens'' scan improves on the trivial scan. The
two curves are $f(\pi)$ and $d(\pi)$. Figure \ref{fig. hmm_diff} shows the actual improvement made by the ``odds-then-evens" scanning and filtering. 

For $\delta=\pi=0.1$, for example, the ``odds-then-evens" error rate is smaller than that of filtering with the trivial scan by $0.021$ (that is, $0.079$ compared to $0.1$). This value should be put alongside the upper bound on the excess loss given in Theorem \ref{theo. non sensitivity to scan in filtering}, which is smaller than $0.07$ in this case. To evaluate the bound on the best achievable scanning and filtering performance given in Theorem \ref{theo. scantering performance} for this example (denoted, with a slight abuse of notation, as $\tilde{U}(\pi,\delta)$), we have
\begin{eqnarray}
\tilde{U}(\pi,\delta) &\ge& \bar{\zeta}^{-1}(\bar{H}(\pi,\delta))
\nonumber\\
& \ge & \bar{\zeta}^{-1}(h_b(\pi * \delta)),
\end{eqnarray}
where $\bar{H}(\pi,\delta)$ is the entropy rate of the output, which is in turn lower bounded by $h_b(\pi * \delta)$. The resulting bound for $\pi=\delta=0.1$ is approximately $0.04$. 

Thus, there exist non-trivial scanning and filtering schemes (i.e., lower bounds) whose improvement on the trivial scanning order is of the same order of magnitude as the upper bound in Theorem \ref{theo. non sensitivity to scan in filtering}. To conclude, it is
clear that there is a wide region were a non trivial scanning order
improves on the trivial scan, an that this region includes at least
all the region between $f(\pi)$ and $d(\pi)$. Yet, it is not clear
what is the optimal scanner-filter pair.
\end{example}
%
%
\subsection{Universal Scanning and Filtering of Noisy Data Arrays}\label{sec. universal scantering}
In \cite{Weiss_et_al07}, Weissman \emph{et}.\ \emph{al}.\ mention that the problems involving sequential decision making on noisy data are not fundamentally different from their noiseless analogue, and in fact can be reduced to the noiseless setting using a properly modified loss function. Indeed, this property of the noisy setting was used throughout the literature, and in this work. The problem of filtering a noisy data sequence is not different in this sense, and it is possible to construct a modified loss function such that the filtering problem is transformed into a prediction problem (with a few important exceptions to be discussed later). Such a modified loss function and a `filtering-prediction transformation' is discussed in \cite{Weiss_et_al07}. We briefly review this transformation, and consider its use in universal filtering of noisy data arrays.

First, we slightly generalize our notion of a filter. For a random variable $U_t$ uniformly distributed on $[0,1]$, let $\hat{X}_t(y^{t-1}y_t,U_t) \in A$ denote the output of the filter $\hat{X}$ at time $t$, after observing $y^t$. That is, the filter $\hat{X}$ also views an auxiliary random variable, on which it can base its output, $\hat{X}_t(y^{t-1}y_t,U_t)$. We also generalize the prediction space to $\calM(\calS)$, $\calS = \{s:N \mapsto A\}$. I.e., the prediction space is a distribution on the set of functions from the noisy observations alphabet $N$ to the clean signal alphabet $A$. We assume an invertible discrete memoryless channel.

For each filter $\hat{X}$, the corresponding predictor is defined by
\begin{equation}
F^{\hat{X}}_t(y^{t-1})[s] = P\left(\hat{X}_t(y^{t-1}y,U_t)=s(y) \quad \forall y \in N \right).
\end{equation}
The analogous `prediction-filtering transformation' is
\begin{equation}
\hat{X}^F_t(y^t,u_t)=a_j \in A \quad \text{if} \quad \sum_{i=0}^{j-1}{\sum_{s:s(y_t)=a_i}{F_t(y^{t-1})[s]}} \leq u_t <\sum_{i=0}^{j}{\sum_{s:s(y_t)=a_i}{F_t(y^{t-1})[s]}},
\end{equation}
where the subscript $i$ reflects some enumeration of $A$. Under the above definitions, \cite[Theorem 4]{Weiss_et_al07} states that for all $n$, $x^n \in A^n$ and any predictor $F$,
\begin{equation}
EL_{\hat{X}^F}(x^n,Y^n)=EL'_F(Y^n),
\end{equation}
where $L_{\hat{X}^F}(x^n,Y^n)$ is the cumulative loss of the filter under the original loss function $l$ and $L'_F(Y^n)$ is the cumulative loss of the predictor under a \emph{modified loss function} $l'$, which depends on the original loss $l$ and the channel crossover probabilities.

This result can be used for universal filtering, under invertible discrete memoryless channels, in the following way. For each finite set of filters, construct the corresponding set of predictors, then use the well known results in universal prediction in order to construct a universal predictor for that set. Finally, construct the universal filter using the ``inverse" prediction-filtering transformation. Analogously, the results on universal finite set scandiction given in \cite{Cohen_Merhav_Weissman_I06} can be used to construct universal scanner-filter pairs. Note, however, that the modified loss function $l'$ may be much more complex to handle compared to the original one. For example, it may not be a function of the difference $x_t-F_t$, even if the original loss function is. Nevertheless, the results in \cite{Cohen_Merhav_Weissman_I06} apply to any bounded loss function, and thus can be utilized.         

\section{Scandiction of Noisy Data Arrays}\label{sec. prediction}
In this section, we consider a scenario similar to that
of Section \ref{sec. filtering}, only now, for
each $t$, the data $Y_{\Psi_t}$ is not available in the estimation of $X_{\Psi_t}$, namely,
$F_t=F_t(Y_{\Psi_1},\ldots,Y_{\Psi_{t-1}})$, as opposed to
$\tilde{F}_t=\tilde{F}_t(Y_{\Psi_1},\ldots,Y_{\Psi_{t}})$ in the filtering scenario. We refer to this scenario as ``noisy scandiction", analogous to the noisy prediction problems discussed in \cite{Weiss_Mer_Somekh01} and \cite{Weiss_Mer01}. 

We first assume the joint probability distribution of the underlying field and noisy observations, $Q$, is known, and examine the settings of Gaussian fields under squared error loss and binary fields under Hamming loss. In these cases, we characterize the noisy scandictability and the achieving scandictors in terms on the ``clean" scandictability of the noisy data. We then consider universal scandiction for the noisy setting, show that this is indeed possible for finite scandictor set and for the class of all stationary binary fields corrupted by binary noise. Finally, we derive bounds on the excess loss when non optimal scanners are used (yet, with the optimal predictor for each scan). 
\subsection{Noisy Scandictability}\label{sec. Prediction
  of Noisy Data Arrays}
Throughout this section, it will be beneficial to consider also the
\emph{clean} scandictability as defined in \cite[Definition
2]{Mer_Weiss03}, that is, when the
scandictor is judged with respect to the same random field it observes.
Thus, for $(X,Y)$ governed by the probability measure $Q$, $Q_Y$ denotes
the marginal measure of $\{Y\}$, and therefore $U(l,Q_Y)$ refers to the clean scandictability
of $Y$, i.e.,
\begin{equation} 
L_{(\Psi,F)}(y_B) =
\sum_{t=1}^{|B|}{l(y_{\Psi_t},F_t(y_{\Psi_1},\ldots,y_{\Psi_{t-1}}))},
\end{equation}
and
\begin{equation} 
U(l,Q_Y) = \lim_{n \rightarrow \infty}\inf_{(\Psi,F)} E_{Q_Y} \frac{1}{|B|}
L_{(\Psi,F)}(Y_B).
\end{equation}
As mentioned earlier, in this section we relate the noisy scandictability, $\bar{U}(l,Q)$, to the clean scandictability of the noisy field, $U(l,Q_Y)$. This relation can be used to derive bounds on $\bar{U}(l,Q)$ using the bounds on $U(l,Q_Y)$ derived in \cite{Mer_Weiss03}.  However, this should be done
carefully. For example, the lower and upper bounds given in
\cite[Theorem 9]{Mer_Weiss03} are applicable only when $X$ has an
autoregressive representation (with respect to some scandictor) with
independent innovations. Unfortunately, $Y=X+N$ does not necessarily
have this representation, and the bounds do not apply to $Y$ in a
straightforward manner.\footnote{Note that the restriction
to autoregressive fields is merely technical, i.e., it facilitates the
proof of the lower bound in the sense that a weak AEP-like theorem
is required. The essence of the lower bound, however, which is a
volume preservation argument, is valid for non autoregressive fields
as well.} Yet, a simple generalization of the lower bound in \cite{Mer_Weiss03}, valid for arbitrarily distributed random fields, can be derived using the same method used in the proof of Theorem \ref{theo. scantering performance}. To this end, we briefly describe this generalization.

Let 
\begin{equation}
B(P) = \min_{\hat{y}} \sum_{y}l(y,\hat{y})P(y),
\end{equation}
and further define
\begin{equation}
\gamma(d) = \max\{H(P) : B(P) \leq d\}.
\end{equation}
Similarly as in Section \ref{sec. Filt. Noisy Data Arrays}, denote by $\bar{\gamma}(d)$ the upper concave envelope of $\gamma(d)$. 
\begin{corollary}\label{cor. lower bound clean scandictability} 
For any random field $Y_B$ and any scandictor $(\Psi,F)$ for $Y_B$,
\begin{equation}
\bar{\gamma}(L_{(\Psi,F)}(Y_B)) \geq \frac{1}{|B|}H(Y_B).
\end{equation}
\end{corollary}
\begin{proof}{}
The proof is similar to that of Theorem \ref{theo. scantering performance}. We have,
\begin{eqnarray}
H(Y_B) & = & H(\Psi(Y_B))
\nonumber\\
&=& \sum_{t=1}^{|B|}{H(Y_{\Psi_t}|Y^{\Psi_{t-1}})}
\nonumber\\
& \leq & \sum_{t=1}^{|B|}{\sum_{y^{\Psi_{t-1}}}\gamma\left(E_{Q_B}\left\{l\left(Y_{\Psi_t},F_{t}(y^{\Psi_{t-1}})\right)|Y^{\Psi_{t-1}}=y^{\Psi_{t-1}}\right\}\right)P(y^{\Psi_{t-1}})}
\nonumber\\
&\leq& |B|\bar{\gamma}\left(\frac{1}{|B|}E_{Q_B}L_{(\Psi,F)}(Y_B)\right).
\end{eqnarray}
\end{proof}
The lower bound in Corollary \ref{cor. lower bound clean scandictability} strengthens the bound in \cite[Theorem 9]{Mer_Weiss03} since it applies to general loss functions, arbitrarily distributed random fields, and is non-asymptotic. When $A = \Reals$ and the loss function is of the form $l(x,F)=\rho(x-F)$, where $\rho(z)$ is monotonically increasing for $z>0$, monotonically decreasing for $z>0$, satisfies $\rho(0)=0$ and $\int{e^{-s\rho(z)}dz}<\infty$ for every $s>0$, the above bound coincides with that of \cite{Mer_Weiss03}. In that case, $\bar{\gamma}(d)=\gamma(d)$, which is in turn the \emph{one sided Fenchel-Legendre transform} of the \emph{log moment generating function} associated with $\rho$ (See \cite[Section III]{Mer_Weiss03} for the details). For example, when $\rho(z)= z^2$, we have, $\gamma(d)=\frac{1}{2}\ln{(2\pi ed)}$, $d>0$ and $\gamma^{-1}(h)=\frac{1}{2\pi e}e^{2h}$, $h>0$. Similar results can be derived for binary alphabet, thus, when $\rho(z)$ is the Hamming loss function, $\gamma(d) = h_b(d)$.  

We now turn to discuss the noisy scnadictability, $\bar{U}(l,Q)$. The following lemma, proved in Appendix \ref{app. proof of lemma scandictability for additive and square error}, describes the noisy scandictability for any additive white noise channel model and the squared error loss function, $l_s(\cdot)$, in terms of the clean scandictability of $Y$, and gives the optimal scandictor.
%
\begin{lemma} \label{lem. scandictability for additive and square error}
Let $\{(X_t,Y_t)\}_{t \in \Z^2}$ be a random field governed by a
probability measure $Q$ such that $Y_t = X_t + N_t$, where $N_t$, $t \in
\Z^2$, are i.i.d.\ random variables with $\text{Var}(N_t) = \sn < \infty$. Then
\begin{equation}
\bar{U}(l_s,Q) = U(l_s,Q_Y) - \sn.
\end{equation}
Furthermore, $\bar{U}(l_s,Q)$ is achieved by the scandictor which
achieves $U(l_s,Q_Y)$.
\end{lemma}
Actually, Lemma \ref{lem. scandictability for additive and square error}
is only scarcely related to scanning. It merely states that in the
prediction of a process based on its noisy observations, under the
additive model stated above and squared error loss, the optimal predictor
is one which disregards the noise, and attempts to predict the next
\emph{noisy} outcome. Similar results for binary processes through
a BSC were given in \cite{Weiss_Mer04} and will be discussed later.

Finally, we mention that the method used in the proof of Lemma \ref{lem.
scandictability for additive and square error} is specific for the square error
loss function. For a general loss function, one can use conditional expectation in order to compute the noisy scandictability, under a \emph{modified} loss function $\rho$. Specifically, for a random field $X$, denote by $\sigma(X_{V_n})$ the smallest sigma algebra
with respect to which $X_{V_n}$ is measurable. Let $\Psi_n$ denote a scanner
for $V_n$ and denote by $\calF^{\Psi_n}_t$ the \emph{information available to
the scandictor at the $t$'th step}, that is 
\begin{equation}
\calF^{\Psi_n}_t =
\sigma\left(Y_{\Psi_1},Y_{\Psi_2(Y_{\Psi_1})},\ldots,Y_{\Psi_t(Y_{\Psi_1}^{\Psi_{t-1}})}\right).
\end{equation}
Note that the set of sites $\Psi_1,\Psi_2,\ldots,\Psi_t$ is itself
random, yet for each $t$, $\Psi_t$ is $\calF^{\Psi_n}_{t-1}$ measurable (if $\Psi$ is random, namely, it uses additional independent random variables, the definition of $\calF^{\Psi_n}_t$ is altered accordingly). Hence,
the filtration $\{\calF^{\Psi_n}_t\}_{t=1}^{|V_n|}$ represents the gathered
knowledge at the scandictor. We have
\begin{eqnarray}
E_{Q_{B_n}}\frac{1}{|B_n|}\sum_{t=1}^{|B_n|}{\rho \left(F_t-Y_{\Psi_t}\right)} &=&
E_{Q_{B_n}} \frac{1}{|B_n|} \sum_{t=1}^{|B_n|}{E_{Q_{B_n}} \left\{ \rho
\left(F_t-X_{\Psi_t}-N_{\Psi_t}\right) \Big | \calF^{\Psi}_{t-1},\sigma(X_{\Psi_t}) \right\}}
\nonumber\\
&=& E_{Q_{B_n}} \frac{1}{|B_n|} \sum_{t=1}^{|B_n|}{ \tilde{\rho}
\left(F_t-X_{\Psi_t}\right)},
\end{eqnarray}
for some $\tilde{\rho}$. Thus, if $l(X_{\Psi_t},F_t)$ is the required loss
function in the noisy prediction problem of $\{X\}$, one has to seek a function
$\rho(\cdot,\cdot)$ such that
$\tilde{\rho}(x_{\Psi_t},F_t)=l(x_{\Psi_t},F_t)$ for all $x_{\Psi_t}$ and $F_t$.
If such a function is found, then surely
$E\rho(Y_{\Psi_t},F_t)=El(X_{\Psi_t},F_t)$ and the optimal scandictor for
the noisy prediction problem is the one which is optimal for the clean
prediction problem of $\{Y\}$ under $\rho$. While this is simple for the squared error loss function and additive noise (choose $\rho(y-F)=(y-F)^2-\sn$), or Hamming loss and BSC (choose $\rho(y,F) = \frac{l_H(y,F)-\delta}{1-2\delta}$) this is not always the case for a general loss function. It is also important to note that in the case of white noise considred in this paper, the condition on the modified loss function $\rho$ can be stated in a single letter expresion, namly, if $l(X,F)$ is the required loss function for the noisy scandiction problem, $\rho$ should satisfy $E\{\rho(Y,F)|\sigma(X)\}=l(X,F)$. 
\subsubsection{Gaussian Random Fields}
Let both $X$ and $N$ be Gaussian random fields,
where the components of $N$ are i.i.d.\ and independent of $X$. That is,
$Y$ is the output of an AWGN channel, with a
Gaussian input $X$. In this scenario, similarly to the clean one, the
noisy scandictability is known exactly and is given by a single letter
expression.

Before we proceed, several definitions are required. For any $t \in \Z^2$
and $V \subseteq \Z^2$, denote by $\hat{X}_t(V)$ the best linear predictor
of $X_t$ given $\{X_{t'}\}_{t' \in V}$. A subset $S \subseteq \Z^2$ is
called a \emph{half plane} if it is closed to addition and satisfies $S
\cup (-S) = \Z^2$ and $S \cap (-S) = \{0\}$. For example,
$S_{\text{lex}}=\{(m,n)\in\Z^2:[m>0] \hspace{1mm} \text{or} \hspace{1mm} [m=0,n \geq 0]\}$ is a half
plane. Let $X$ be a wide sense
stationary random field and denote by $g$ the density function associated
with the absolutely continuous component in the Lebesgue decomposition of
its spectral measure. Then, for any half plane $S$, we have \cite[Theorem
1]{Helson_Lowdenslager58},
\begin{eqnarray} \label{def. sigma_x}
E\left(X_0-\hat{X}_0(-S \setminus \{0\})\right)^2 &=&
\exp\left\{\frac{1}{4\pi^2}\int_{[0,2\pi)^2}{\ln
g(\lambda)d\lambda}\right\}
\nonumber\\
&\defined& \sigma_u^2(X).
\end{eqnarray}
We can now state the following corollary, regarding the noisy
scandictability in the Gaussian regime and squared error loss, which is a
direct application of Lemma \ref{lem. scandictability for additive and
square error} and the results of \cite[Section IV]{Mer_Weiss03}.
%
\begin{corollary} \label{coro. noisy scand. gaussian}
Let $\{(X_t,Y_t)\}_{t \in \Z^2}$ be a random field governed by a
probability measure $Q$ such that $Y_t = X_t + N_t$, where $X$ is a stationary
Gaussian random field, $N_t$, $t \in \Z^2$, is an AWGN, independent of $\{X_t\}_{t \in \Z^2}$. Then, the noisy
scandictability of $Q$ under the squared error loss is given by
\begin{equation} \label{eq. noisy scan. gaussian - as sigma}
\bar{U}(l_s,Q) = \sigma_u^2(Y) - \sn.
\end{equation}
Furthermore, $\bar{U}(l_s,Q)$ is asymptotically achieved by a scandictor which
scans $(X_t,Y_t)$ according to the total order defined by any half-plane
$S$ and applies the corresponding best linear predictor for the next
outcome of $Y$.
\end{corollary}
For any stationary Gaussian \emph{process} $X$, it has been shown by
Kolmogorov (see for example \cite{Papoulis84}) that the entropy rate is
given by
\begin{equation}\label{def. ent rate}
\hsx = \frac{1}{2}\ln{(2\pi e)} +
\frac{1}{4\pi}\int_{-\pi}^{\pi}{\ln{g(\lambda)}d\lambda}.
\end{equation}
Thus, using the one-dimensional analogue of \eqref{def. sigma_x}, for a
stationary Gaussian process $X$ we have,
\begin{equation} \label{eq. hsx as a function of sx}
\hsx = \frac{1}{2}\ln{(2\pi e \sigma_u^2(X))}.
\end{equation}
In fact, \eqref{eq. hsx as a function of sx} applies for stationary
Gaussian random fields as well. Thus, we have,
\begin{eqnarray}
\bar{U}(l_s,Q) &=& \sigma_u^2(Y) - \sn
\nonumber\\
& = & \frac{1}{2\pi e} e^{2 \hsy} - \frac{1}{2\pi e} e^{2 H^N}, \label{eq.
noisy scan. gaussian - as entropy}
\end{eqnarray}
where $\hsy$ is the entropy rate of $Y$ and $H^N$ is the entropy of each
$N_t$. From the entropy power inequality \cite[pp. 496]{Cov_Thom91}, we
have,
\begin{equation} \label{eq. entropy power inequality}
\frac{1}{2\pi e}e^{2\hsy} \geq \frac{1}{2\pi e}e^{2 H^N }+\frac{1}{2\pi
e}e^{2\hsx},
\end{equation}
thus, as expected, the noisy scandictability given in Corollary \ref{coro.
noisy scand. gaussian} (and \eqref{eq. noisy scan. gaussian - as entropy})
is at least as small the clean scandictability of $X$, that is, with no
noise at all. In most of the interesting cases, however, \eqref{eq.
entropy power inequality} is a strict inequality. In fact, as mentioned in
\cite{Blachman65}, \eqref{eq. entropy power inequality} is achieved with
equality only when both $X$ and $N$ are Gaussian and have
\emph{proportional spectra}. Consequently, unless $X$ is white, Corollary
\ref{coro. noisy scand. gaussian} is non-trivial.
\subsubsection{Binary Random Fields}
In this case, the
results of \cite{Weiss_Mer_Somekh01} and \cite{Weiss_Mer04}
shed light on the optimal scandictor. Therein,
it was shown that for a binary prediction problem, i.e., where $\{X_t\}$
is a binary source passed through a BSC with
cross over probability $\delta<\frac{1}{2}$, and $\{Y_t\}$
is the channel output, the more likely outcome for the clean bit is
also the more likely outcome for the noisy bit. Thus, the optimal
predictor in the Hamming sense for the next clean bit (based on the
noisy observations) might as well use the same strategy
as if it tries to predict the next noisy bit. Consequently, the
optimal \emph{scandictor} in the noisy setting is the one
which is optimal for $\{Y\}$, and the results of \cite[Section
V]{Mer_Weiss03} apply.

The following proposition relates the scandictability of a binary
noise-corrupted process $\{Y\}$, judged with respect to the clean binary
process $\{X\}$, to its clean scandictability.
\begin{proposition}\label{prop. noisy scndiction binary}
Let $\{(X_t,Y_t)\}_{t \in \Z^2}$ be a binary random field governed by
a probability measure $Q$ such that $\{Y_t\}$ is the output of a
binary memoryless symmetric channel with cross over probability
$\delta$ and input $\{X_t\}$. Then,
\begin{equation}
\bar{U}(l_H,Q)=\frac{U(l_H,Q_Y)-\delta}{1-2\delta},
\end{equation}
where $l_H$ is the Hamming loss function. Furthermore, $\bar{U}(l_H,Q)$ is achieved by the scandictor which
achieves $U(l_H,Q_Y)$.
\end{proposition}
Note that indeed $U(l_H,Q_Y) \geq \delta$ as $Y$ is the output of a BSC with crossover probability $\delta$.
\begin{proof}[Proof (Proposition \ref{prop. noisy scndiction binary})]
Let $\{B_n\}_{n \geq 1}$ be any sequence of elements in $\calV$,
satisfying $R(B_n) \rightarrow \infty$. We have,
\begin{eqnarray}
\bar{U}(l_H,Q_{B_n})&=&\inf_{(\Psi,F) \in \mathcal{S}(B_n)}
E_{Q_{B_n}}
\frac{1}{|B_n|}\sum_{t=1}^{|B_n|}{l_H(X_{\Psi_t},F_t(Y_{\Psi_1},\ldots,Y_{\Psi_{t-1}}))}
\nonumber\\
&=&\inf_{(\Psi,F) \in \mathcal{S}(B_n)}
\frac{1}{|B_n|}\sum_{t=1}^{|B_n|}{P\left(F_t(Y_{\Psi_1},\ldots,Y_{\Psi_{t-1}})
      \ne X_{\Psi_t}\right)}, \label{eq. barU in terms of px}
\end{eqnarray}
and, analogously,
\begin{equation}
U(l_H,Q_{Y,B_n})=\inf_{(\Psi,F) \in \mathcal{S}(B_n)}
\frac{1}{|B_n|}\sum_{t=1}^{|B_n|}{P\left(F_t(Y_{\Psi_1},\ldots,Y_{\Psi_{t-1}})
      \ne Y_{\Psi_t}\right)}.
\end{equation}
Denoting by $Z_t$ the channel noise at time $t$, and abbreviating
$F_t(Y_{\Psi_1},\ldots,Y_{\Psi_{t-1}})$ by $F_t$, we have
\begin{eqnarray}
P\left(F_t \ne Y_{\Psi_t}\right) &=& P\left(F_t \ne Y_{\Psi_t},
  Z_{\Psi_t}=1\right) + P\left(F_t \ne Y_{\Psi_t}, Z_{\Psi_t}=0\right)
\nonumber\\
&=& P\left(F_t = X_{\Psi_t},
  Z_{\Psi_t}=1\right) + P\left(F_t \ne X_{\Psi_t}, Z_{\Psi_t}=0\right)
\nonumber\\
&=& \left(1-P\left(F_t \ne X_{\Psi_t}\right)\right)\delta + P\left(F_t \ne X_{\Psi_t}\right)(1-\delta).
\end{eqnarray}
Namely, for $\delta<\frac{1}{2}$, the optimal strategy for
predicting $Y_{\Psi_t}$ based on $Y_{\Psi_1},\ldots,Y_{\Psi_{t-1}}$
and the optimal strategy for
predicting $X_{\Psi_t}$ based on $Y_{\Psi_1},\ldots,Y_{\Psi_{t-1}}$
are identical, and, in addition,
\begin{equation}\label{eq. relation px py}
P\left(F_t \ne X_{\Psi_t}\right)=\frac{P\left(F_t \ne
    Y_{\Psi_t}\right)-\delta}{1-2\delta}.
\end{equation}
Substituting \eqref{eq. relation px py} into \eqref{eq. barU in terms of
px} and taking $n \to \infty$ completes the proof.
\end{proof}
%
\subsection{Universal Scandiction in the Noisy
 Scenario}\label{sec. universal scandiction noisy}Section \ref{sec. Prediction of Noisy Data
 Arrays} dealt with the actual
value of the best achievable performance in the noisy scandiction scenario. However,
it is also interesting to investigate the universal setting in which one seeks a
predictor which does not depend on the joint probability measure of
$\{(X,Y)\}$, yet performs asymptotically as well as a one matched to this
measure. The problem of universal scandiction in the noiseless scenario was dealt with in \cite{Cohen_Merhav_Weissman_I06}. Herein, we show that it is possible to construct universal scandictors in the noisy setting as well (similar to universal scanning and filtering in Section \ref{sec. universal scantering}). First, we show that it is possible to compete successfully with any finite set of scandictors, and present a universal scandictor for this setting. We then show that with a proper choice of a set of scandictors, it is possible to (universally) achieve $\bar{U}(l,Q)$, i.e., the noisy scandictability, for any spatially stationary random filed $(X,Y)$. 

At the basis of the results of \cite{Cohen_Merhav_Weissman_I06} stands the exponential weighting algorithm, originally derived by Vovk in \cite{Vovk90}. In \cite{Vovk90}, Vovk considered a general set of \emph{experts} and introduced the exponential weighting
algorithm in order to compete with the best expert in the set. In this algorithm, each expert is assigned with a weight,
according to its past performance. By decreasing the weight of poorly
performing experts, hence preferring the ones proved to perform well thus
far, one is able to compete with the best expert, having neither any
\emph{a priori} knowledge on the input sequence nor which expert will
perform the
best. It is clear that the essence of this algorithm is the use of the cumulative losses incurred by each expert to construct a probability measure on the experts, which is later used to choose an expert for the next action. However, when the clean data $X$ is not known to the sequential algorithm, it is impossible to calculate the cumulative losses of the experts precisely. Nevertheless, as Weissman and Merhav show in \cite{Weiss_Mer01}, using an unbiased estimate $\hat{X_t}(Y_t)$ of $X_t$ results in sufficiently accurate estimates of the cumulative losses of the experts, which in turn can be used by the exponential weighting algorithm. Hence, the framework derived in \cite{Cohen_Merhav_Weissman_I06} can then be used to suggest universal scandictors for the noisy setting as well. 

Consider a random field $(X_B,Y_B)$ where $X$ is binary and $Y$ is either binary (e.g., the output of a BSC whose input is $Y$) or real valued (e.g., $X$ through a Gaussian noise channel). For a loss function $l:\{0,1\}\times[0,1] \to [0,\infty]$ we define, similarly to \cite{Weiss_Mer01},
\begin{equation}
l_0(\cdot) \defined l(0,\cdot) \quad \text{and} \quad l_1(\cdot) \defined l(1,\cdot).
\end{equation}
Assume $(\Psi,F)$ is a scandictor for $B$. Then, for any $t \leq |B|$, we have
\begin{eqnarray}
L_{(\Psi,F)}(x_B,y_B)_t&=&\sum_{i=1}^{t}l(F_i(y_{\Psi_1}^{\Psi_{i-1}}),x_{\Psi_i})
\nonumber\\
&=&\sum_{i=1}^{t}\left[(1-x_{\Psi_i})l_0(F_i(y_{\Psi_1}^{\Psi_{i-1}}))+x_{\Psi_i}l_1(F_i(y_{\Psi_1}^{\Psi_{i-1}}))\right].
\end{eqnarray}
Clearly, $L_{(\Psi,F)}(x_B,y_B)_t$ depends on $x_B$ and is not known to the sequential algorithm. Let $h(y_{\Psi_i})$ be an unbiased estimate for $x_{\Psi_i}$. For example, when $Y$ is the output of a BSC with input $X$ we may choose 
\begin{equation}
h(y_{\Psi_i})=\frac{y_{\Psi_i}-\delta}{1-2\delta}. 
\end{equation}
Define
\begin{equation}
\hat{L}_{(\Psi,F)}(y_B)_t=\sum_{i=1}^{t}\left[(1-h(y_{\Psi_i}))l_0(F_i(y_{\Psi_1}^{\Psi_{i-1}}))+h(y_{\Psi_i})l_1(F_i(y_{\Psi_1}^{\Psi_{i-1}}))\right],
\end{equation}
and
\begin{eqnarray}\label{def. Delta}
\Delta_{(\Psi,F)}(x_B,y_B)_t &\defined& L_{(\Psi,F)}(x_B,y_B)_t-\hat{L}_{(\Psi,F)}(y_B)_t
\nonumber\\
&=&\sum_{i=1}^{t}\left(h(y_{\Psi_i})-x_{\Psi_i}\right)l_0(F_i(y_{\Psi_1}^{\Psi_{i-1}}))+\sum_{i=1}^{t}\left(x_{\Psi_i}-h(y_{\Psi_i})\right)l_1(F_i(y_{\Psi_1}^{\Psi_{i-1}})).
\nonumber\\
\end{eqnarray}
Similarly to \cite{Weiss_Mer01}, we assume that the noise field $N_B$ is of
independent components and that for each $i \in B$, $Y_i \in \sigma(N_i)$,
i.e., the noise component at site $i$ affects the observation at that site
alone. In Appendix \ref{app. proof of martingale}, we show that for any
image $x_B$ and any scandictor $(\Psi,F)$ for $B$,
$\left(\Delta_{(\Psi,F)}(x_B,y_B)_t,\calF^\Psi_t\right)$ is a zero mean martingale. As a result, for any scandictor $(\Psi,F)$, image $x_B$ and $t$ we have
\begin{equation}\label{eq. equality in expectation of L and hatL}
E L_{(\Psi,F)}(x_B,Y_B)_t = E \hat{L}_{(\Psi,F)}(Y_B)_t,
\end{equation}
namely, $\hat{L}_{(\Psi,F)}(Y_B)_t$ is an unbiased estimator for
$L_{(\Psi,F)}(x_B,Y_B)_t$. The universal algorithm for scanning and prediction
in the noisy scenario will thus use $\hat{L}_{(\Psi,F)}(Y_B)_t$ instead of
$L_{(\Psi,F)}(x_B,Y_B)_t$, which is unknown. More specifically, similarly to
the algorithm proposed in \cite{Cohen_Merhav_Weissman_I06}, the
algorithm divides the data array to be scandicted to blocks of size $m(n)\times
m(n)$, then scans the data in a (fixed) block-wise order, where each block is
scandicted using a scandictor chosen at random from the scandictors set,
according to the distribution
$\hat{P}_i\left(j|\{\hat{L}_{j,i}\}_{j=1}^{\lambda}\right)$,
\begin{equation}
\hat{P}_i\left(j|\{\hat{L}_{j,i}\}_{j=1}^{\lambda}\right)=\frac{e^{-\eta
\hat{L}_{j,i}}}{\sum_{j=1}^{\lambda}{e^{-\eta \hat{L}_{j,i}}}},
\end{equation}
where $\hat{L}_{j,i}=\sum_{m=0}^{i-1}\hat{L}_{(\Psi,F)_j}(y^m)$, the estimated cumulative loss of the scandictor $(\Psi,F)_j$ after scandicting $i$ blocks of data, when $(\Psi,F)_j$ is restarted after each block, and $\lambda$ is the cardinality of the set of scandictors, $\calF_m$.\footnote{To be consistent with the notation of \cite{Cohen_Merhav_Weissman_I06}, the same notation is used for both a filtration and a scandictor set. The difference should be clear from the context.} Note the subscript $m$ in $\calF_m$, as in order to scandict a data array of size $n \times n$, the universal algorithm discussed herein uses the scandictors with which it competes, but only on blocks of size $m \times m$.

The following proposition gives an upper bound on the redundancy of the algorithm when competing with a finite set of scandictors, each operating block-wise on the data array.
\begin{proposition}\label{prop. compete with a set of scan. block-wise, noisy}
Let $EL_{alg}(x_{V_n},Y_{V_n})$ be the expected (with respect to the noisy
random field as well as the randomization in the algorithm) cumulative loss of
the proposed algorithm on $Y_{V_n}$, when the underlying clean array is
$x_{V_n}$ and the noisy field is of independent components with $Y_i \in
\sigma(N_i)$ for each $i \in V_n \subset \Z^2$. Let $EL_{min}(x_{V_n},Y_{V_n})$
denote the expected cumulative loss of the best scandictor in $\calF_m$,
operating block-wise on $Y_{V_n}$. Assume $|\calF_m|=\lambda$, then
\begin{equation} \label{eq. regret in compete with a set of scan. block-wise, noisy}
EL_{alg}(x_{V_n},Y_{V_n}) - EL_{min}(x_{V_n},Y_{V_n}) \leq
m(n)(n+m(n))\sqrt{\ln \lambda}\frac{l_{max}}{\sqrt{2}}.
\end{equation}
\end{proposition}
\begin{proof}{}
By \eqref{eq. equality in expectation of L and hatL} and \cite[Proposition 3]{Cohen_Merhav_Weissman_I06}, for any $x_{V_n}$ we have
\begin{align}
EL_{alg}(x_{V_n},Y_{V_n}) - \min_{(\Psi,F) \in \calF_m}&
EL_{(\Psi,F)}(x_{V_n},Y_{V_n})
\nonumber\\
&= E\hat{L}_{alg}(Y_{V_n}) - \min_{(\Psi,F) \in \calF_m}
E\hat{L}_{(\Psi,F)}(Y_{V_n})
\nonumber\\
&\leq E\hat{L}_{alg}(Y_{V_n}) -  E\min_{(\Psi,F) \in
\calF_m}\hat{L}_{(\Psi,F)}(Y_{V_n})
\nonumber\\
&= E\left\{\hat{L}_{alg}(Y_{V_n}) -  \min_{(\Psi,F) \in
\calF_m}\hat{L}_{(\Psi,F)}(Y_{V_n})\right\}
\nonumber\\
&\leq m(n)(n+m(n))\sqrt{\ln \lambda}\frac{l_{max}}{\sqrt{2}}.
\end{align}
\end{proof}
Proposition \ref{prop. compete with a set of scan. block-wise, noisy} is the basis for the main result in this sub-section, a universal scandictor which competes successfully with any finite set of scandictors for the noisy scenario.
\begin{theorem} \label{theo. universal finite set scandictability, noisy}
Let $(X,Y)$ be a stationary random field with a probability measure $Q$. Assume that for each $i \in \Z^2$, $Y_i$ is the output of a memoryless channel whose input is $X_i$. Let
$\calF = \{\calF_n\}$ be an arbitrary sequence of scandictor sets, where $\calF_n$ is a set of scandictors for $V_n$ and $|\calF_n| = \lambda < \infty$ for all $n$. Then, there
exists a sequence of scandictors
$\{(\hat{\Psi},\hat{F})_n\}$, independent of $Q$, for which
\begin{equation} \label{eq. inq. between the liminf, noisy}
\liminf_{n \rightarrow \infty}
E_{Q_{V_n}}E\frac{1}{|V_n|}L_{(\hat{\Psi},\hat{F})_n}(X_{V_n},Y_{V_n}) \leq
\liminf_{n \rightarrow \infty}\min_{(\Psi,F) \in \calF_n}
E_{Q_{V_n}}\frac{1}{|V_n|}L_{(\Psi,F)}(X_{V_n},Y_{V_n})
\end{equation}
for any $Q \in \calM_S(\Omega)$, where the inner expectation in the
l.h.s.\ of \eqref{eq. inq. between the
liminf, noisy} is due to the possible randomization in
$(\hat{\Psi},\hat{F})_n$.
\end{theorem}
The proof of Theorem \ref{theo. universal finite set scandictability, noisy} follows the proof of \cite[Theorem 2]{Cohen_Merhav_Weissman_I06} verbatim.

It is now possible to show the existence of a universal scandictor for any stationary random field in the noisy scandiction setting. Herein, we include only the setting where $X$ is binary and $Y$ is the output of a BSC. In this case, the scandictor is twofold-universal, namely, it does not depend on the channel crossover probability either. Extending the results to real-valued noise is possible using the methods introduced in \cite{Weiss_Mer04} (although the universal predictor does depend on the channel characteristics) and will be discussed later.
%
\begin{theorem}\label{theo. existence of an alg achieving tildeU}
Let $X$ be a stationary random field over a finite alphabet $A$ and a probability measure $Q$. Let $Y$ be the output of a BSC whose input is $X$ and whose crossover probability $\delta$. Let the prediction space $D$ be either finite or bounded (with $l(x,F)$ then being Lipschitz in its second argument). Then, there
exists a sequence of scandictors
$\{(\Psi,F)_n\}$, independent of $Q$ and of $\delta$, for which
\begin{equation}\label{eq. achieving tildeU}
\lim_{n \rightarrow \infty}
E_{Q_{V_n}}E\frac{1}{|V_n|}L_{(\Psi,F)_n}(X_{V_n},Y_{V_n}) = \tilde{U}(l,Q)
\end{equation}
for any $Q \in \calM_S(\Omega)$, where the inner expectation in the
l.h.s.\ of \eqref{eq. achieving tildeU} is due to the possible randomization in
$(\Psi,F)_n$.
\end{theorem}
Similar to \cite[Section A]{Weiss_Mer04} and the proof of \cite[Theorem 6]{Cohen_Merhav_Weissman_I06}, in the case of binary input and
binary-valued noise it is possible to take the set of scandictors with which we
compete as the set of \emph{all possible scandictors for an $m(n) \times m(n)$}
block. The proof thus follows directly from the proof of \cite[Theorem 6]{Cohen_Merhav_Weissman_I06}.

As for continuous-valued observations, it is quite clear that the set of all
possible scandictors for an $m(n)\times m(n)$ block is far too rich to compete
with (note that this is since the number of \emph{predictors} is too large). A
complete discussion is available in \cite[Section B]{Weiss_Mer04}. However,
Weissman and Merhav do offer a method for successfully achieving the Bayes
envelope for this setting, by introducing a much smaller set of predictors,
which on one hand \emph{includes} the best $k$th order Markov predictor, yet on
the other hand is not too rich, in the sense that the redundancy of the
exponential weighting algorithm tends to zero when competing \emph{with an
$\epsilon$-grid of it}. Since presenting a universal \emph{scandictor} for this
scenario will mainly include a repetition of the many details discussed in
\cite{Weiss_Mer04}, we do not include it here.
%
%
%
%
%
%
%
%
\subsection{Bounds on the Excess Loss for Non-Optimal Scandictors}\label{sec. bounds scandiction}
Analogously to the scanning and filtering setting discussed in Section \ref{sec. filtering}, and the clean prediction setting discussed in \cite{Cohen_Merhav_Weissman_I06}, it is interesting to investigate the excess loss incurred when non optimal scandictors are used in the noisy scandiction setting. Unlike the scanning and filtering setting, where the excess loss bounds were not a straightforward extensions of the results in \cite{Cohen_Merhav_Weissman_I06}, in the noisy scandiction scenario this problem can be quite easily tackled using the results of \cite{Cohen_Merhav_Weissman_I06} and modified loss functions. 

We briefly state the results of \cite{Cohen_Merhav_Weissman_I06} in this context. The scenario considered therein is that of predicting the next outcome
of a binary source, with $D = [0,1]$ as the prediction space. $\phi_\rho$ denotes the Bayes envelope associated with the loss function $\rho$, i.e.,
\begin{equation}\label{def. phi - bayes env}
\phi_\rho(p)=\min_{q\in[0,1]}[(1-p) \rho(0,q)+p \rho(1,q)].
\end{equation}
Similarly to \eqref{def. epsilon_delta}, define
\begin{equation}\label{def. epsilon_l}
\epsilon_\rho=\min_{\alpha,\beta}\max_{0\leq p\leq 1}|\alpha h_b(p)+\beta
- \phi_\rho(p)|.
\end{equation}
Note that although the definitions of $\phi_\rho(p)$ and $\epsilon_\rho$ refer to the binary scenario, the result below holds for larger alphabets, with $\epsilon_\rho$ defined as in \eqref{def. epsilon_l}, with the maximum ranging over the simplex of all distributions on the alphabet, and $h(p)$ (replacing $h_b(p)$) and $\phi_\rho(p)$ denoting the entropy and Bayes envelope of the distribution $p$, respectively. In \cite{Cohen_Merhav_Weissman_I06}, it is shown that if $X_B$ is an arbitrarily distributed binary random field, then, for any scan $\Psi$,
\begin{equation}\label{eq. bound of prop. abs diff with scanning}
\left|\alpha_\rho \frac{1}{|B|}H(X_B)+\beta_\rho-E_{Q_B}\frac{1}{|B|}L_{(\Psi,F^{opt})}(X_B)\right| \leq \epsilon_\rho,
\end{equation}
where $\alpha_\rho$ and $\beta_\rho$ are the achievers of the minimum in
\eqref{def. epsilon_l}.

As mentioned earlier, if
$\rho(Y,F)$ is some loss function for the ``clean" prediction
problem of $\{Y\}$, the noisy process, then,
\begin{equation}
E\left\{\rho(Y,F)|\sigma(X)\right\}=\tilde{\rho}(X,F)
\end{equation}
for some $\tilde{\rho}$. Assuming a suitable $\rho$ is found (i.e., $\tilde{\rho} =l$), we have, for any scan $\Psi$,
\begin{align}
\Bigg|\alpha_\rho\frac{1}{|B|}H(Y_B) & +\beta_\rho-  \frac{1}{|B|}E_{Q_B}L^l_{\Psi,F^{opt}}(X_B,Y_B)\Bigg|
\nonumber\\
&\hspace{+1cm}=\left|\alpha_\rho\frac{1}{|B|}H(Y_B)+\beta_\rho- \frac{1}{|B|}E_{Q_B}L^\rho_{\Psi,F^{opt}}(Y_B)\right|
\nonumber\\
&\hspace{+1cm}\leq \epsilon_\rho,
\end{align}
where $\frac{1}{|B|}E_{Q_B}L^l_{\Psi,F^{opt}}(X_B,Y_B)$ is the normalized expected cumulative loss in optimally predicting $X_{\Psi_t}$ based on $Y_1^{\Psi_{t-1}}$, under the loss function $l$, $\frac{1}{|B|}E_{Q_B}L^\rho_{\Psi,F^{opt}}(Y_B)$ is the normalized expected cumulative loss in optimally predicting $Y_{\Psi_t}$ based on $Y_1^{\Psi_{t-1}}$, under the loss function $\rho$, and $\alpha_\rho$ and $\beta_\rho$ are the minimizers of $\epsilon_\rho$ as defined in \eqref{def. epsilon_l}. Hence, the following corollary applies.
\begin{corollary}\label{cor. bound on excess loss in noisy prediction}
Let $X_B$ be an arbitrarily distributed binary field. Assume a white noise, and denote the noisy version of $X_B$ by $Y_B$. Let $D = [0,1]$ be the prediction space and $l:\{0,1\}\times D \to \Reals$ be any loss function. Then, for any scan $\Psi$,
\begin{equation}
\left|\frac{1}{|B|}E_{Q_B}L^l_{\Psi,F^{opt}}(X_B,Y_B) -\bar{U}(l,Q_B)\right| \leq 2\epsilon_\rho,
\end{equation}
when $\rho$ is a loss function such that
$E\left\{\rho(Y,F)|\sigma(X)\right\}=l(X,F)$ for any $F$.
\end{corollary}
\begin{example}[\textit{BSC and Hamming Loss}]
In the case of binary input, BSC with crossover probability $\delta$ and Hamming loss $l_H(\cdot,\cdot)$, it is not hard to show that
\begin{equation}
\rho(y,F)=\frac{l_H(y,F)-\delta}{1-2\delta}.
\end{equation}
Hence, 
\begin{equation}
\phi_\rho(p) = \frac{\phi_{l_H}(p)-\delta}{1-2\delta}
\end{equation}
and 
\begin{equation}
\epsilon_\rho = \frac{1}{1-2\delta}\epsilon_{l_H},
\end{equation}
where $\epsilon_{l_H}=0.08$ as mentioned in \cite{Cohen_Merhav_Weissman_I06}. 
The above bound on the excess loss can also be computed directly, without using Corollary \ref{cor. bound on excess loss in noisy prediction}, as for any scan $\Psi$, the normalized cumulative prediction errors are given by 
\begin{equation}
\frac{1}{|B|}E_{Q_B}L^{l_H}_{\Psi,F}(X_B,Y_B)=
\frac{1}{|B|}\sum_{t=1}^{|B|}{P\left(F_t(Y_{\Psi_1},\ldots,Y_{\Psi_{t-1}})
      \ne X_{\Psi_t}\right)}
\end{equation}
for the noisy scenario, and
\begin{equation}
\frac{1}{|B|}E_{Q_B}L^{l_H}_{\Psi,F}(Y_B)=
\frac{1}{|B|}\sum_{t=1}^{|B|}{P\left(F_t(Y_{\Psi_1},\ldots,Y_{\Psi_{t-1}})
      \ne Y_{\Psi_t}\right)}
\end{equation}
for the (clean) prediction of $Y_B$. Hence, using \eqref{eq. relation px py}, for any scan $\Psi$ we have,
\begin{align}
&\left|\frac{1}{|B|}E_{Q_B}L^{l_H}_{\Psi,F^{opt}}(X_B,Y_B) -\bar{U}(l_H,Q_{B}) \right|
\nonumber\\
&\hspace{+1cm}= \left| \frac{\frac{1}{|B|}E_{Q_B}L^{l_H}_{\Psi,F^{opt}}(Y_B)-\delta}{1-2\delta} -
  \frac{U(l_H,Q_{Y,B})-\delta}{1-2\delta} \right|
\nonumber\\
&\hspace{+1cm}= \frac{1}{1-2\delta}\left| \frac{1}{|B|}E_{Q_B}L^{l_H}_{\Psi,F^{opt}}(Y_B)-U(l_H,Q_{Y,B}) \right|
\nonumber\\
&\hspace{+1cm}\leq \frac{2\epsilon_{l_H}}{1-2\delta}.
\end{align}
\end{example}
\begin{example}[\textit{Additive Noise and Squared Error}]
Let $Y_B$ be the output of an additive channel, with $\sigma_v^2$ denoting the noise variance. Let $l_s$ be the squared error loss function. In this case,
\begin{equation}
E\Big\{ \underbrace{(Y_{\Psi_t}-F_t(y^{\Psi_{t-1}}))^2-\sigma_v^2}_{\rho(Y_{\Psi_t},F_t(y^{\Psi_{t-1}}))}
|\sigma(X_{\Psi_t}) \Big\} = \underbrace{(X_{\Psi_t}-F_t(y^{\Psi_{t-1}}))^2}_{l_s(X_{\Psi_t},F_t(y^{\Psi_{t-1}}))}.
\end{equation}
Thus, Corollary \ref{cor. bound on excess loss in noisy prediction} applies with $\rho(Y,\hat{Y})=(Y-\hat{Y})^2-\sigma_v^2$, and clearly $\epsilon_\rho = \epsilon_{l_s}$. Note that although Corollary \ref{cor. bound on excess loss in noisy prediction} is stated for binary alphabet, it is not hard to generalize its result to larger alphabets, as mentioned in \cite[Section 4]{Cohen_Merhav_Weissman_I06}.
\end{example}
%
\subsubsection{Excess Loss Bounds Via the Continuous Time Mutual Information}
The bound on the excess noisy scandiction loss given in Corollary \ref{cor. bound on excess loss in noisy prediction} was derived using the results of \cite{Cohen_Merhav_Weissman_I06} and modified loss functions. However, new bounds can also be derived using the same method which was used in the proof of Theorem \ref{theo. bound on sens. of scantering Gaussian}, namely, the scan invariance of the mutual information and the relation to the continuous time problem. We briefly discuss how such a bound can be derived for noisy scandiction of Gaussian fields corrupted by Gaussian noise.

Using the notation of Section \ref{sec. bounds filtering Gaussian}, we have
\begin{eqnarray}\label{def. g}
\Var(X) - \int_0^1 \Var(X_t|Y^t) \der t &=& \sx - \sn \ln\left(1+\frac{\sx}{\sn}\right) 
\nonumber\\
&=& \sn g\left(\frac{\sx}{\sn}\right),
\end{eqnarray}
where
\begin{equation}\label{def. g scandiction}
g(x)=x-\ln(1+x).
\end{equation}
Since $\sx \ge \Var\left(X_{\Psi_i}|Y_{\Psi_1}^{\Psi_{i-1}}\right)$ and $g(x)$ is monotonically increasing for $x >0$, derivations similar to \eqref{eq. proof of scantering bound G} lead to 
\begin{equation}
\frac{1}{n^2}EL_{(\Psi,F^{opt})}\left(X_{V_n},Y_{V_n}\right) \leq \sn g\left(\frac{\sx}{\sn}\right) + \frac{1}{n^2}2 \sn I\left(X_{V_n},Y_{V_n}\right). 
\end{equation}
On the other hand, since $g(x) \ge 0$ for $x \ge 0$, we have
\begin{equation}
\frac{1}{n^2}EL_{(\Psi,F^{opt})}\left(X_{V_n},Y_{V_n}\right) \ge \frac{1}{n^2}2 \sn I\left(X_{V_n},Y_{V_n}\right), 
\end{equation}
which now can be viewed as the scanning and prediction analogue of \cite[eq. (156b)]{Guo_Shamai_Verdu05}. We thus have the following corollary.
\begin{corollary}\label{cor. bound on sens. of scandiction Gaussian}
Let $X_{V_n}$ be a Gaussian random field with a constant marginal distribution satisfying $\Var(X_i)=\sx < \infty$ for all $i \in V_n$. Let $Y_i=X_i+N_i$, where $N_{V_n}$ is a white Gaussian noise of variance $\sigma_N^2$, independent of $X_{V_n}$. Then, for any two scans $\Psi^1$ and $\Psi^2$ and their optimal predictors, we have
\begin{equation}
\frac{1}{n^2}\left|EL_{(\Psi^1,F^{opt})}\left(X_{V_n},Y_{V_n}\right)-EL_{(\Psi^2,F^{opt})}\left(X_{V_n},Y_{V_n}\right)\right| 
\leq \sn g\left(\frac{\sx}{\sn}\right). 
\end{equation}
\end{corollary}
Similarly as in Theorem \ref{theo. bound on sens. of scantering Gaussian}, the bound in Corollary \ref{cor. bound on sens. of scandiction Gaussian} has the form $\sx \frac{g(\snr)}{\snr}$, namely, it scales with the variance of the input. As expected, at the limit of low SNR, $\frac{g(\snr)}{\snr} \to 0$, since regardless of the scan, one is clueless about the underlying clean symbol. In fact, it is not surprising that this behavior is common to both the filtering and the prediction scenarios. In the former, the bound value is given by \eqref{def. f}, while in the latter it is given by \eqref{def. g}. In both cases, the bound value is simply the difference between a continuous time filtering problem, and a discrete time filtering (or prediction, in \eqref{def. g}) problem. It is not hard to see that this difference tends to $0$ as $\snr \to 0^+$. At the limit of high SNR, $\frac{g(\snr)}{\snr} \to 1$. Indeed, this limit corresponds to the noiseless scandiction scenario, where scanning is consequential \cite{Mer_Weiss03}. 

\section{Conclusion}\label{sec. conc}
We investigated problems in sequential filtering and prediction of noisy multidimensional data arrays. A bound on the best achievable scanning and filtering performance was derived, and the excess loss incurred when non-optimal scanners are used was quantified. In the prediction setting, a relation of the best achievable performance to that of the clean scandictability was given. In both the filtering and prediction scenarios, a special emphasis was given to the cases of AWGN and squared error loss, and BSC and Hamming loss.  

Due to their sequential nature, the problems discussed in this paper are strongly related to the filtering and prediction problems where reordering of the data is not allowed (or where there is only one natural order to scan the data), such as robust filtering and universal prediction discussed in the current literature. However, the numerous scanning possibilities in the multidimensional setting add a multitude of new challenges. In fact, many interesting problems remain open. It is clear that identifying the optimal scanning methods in the widely used input and channel models discussed herein is required, as the implementation of universal algorithms might be too complex in realistic situations. Moreover, tighter upper bounds on the excess loss can be derived in order to better understand the trade-offs between non-trivial scanning methods and the overall performance. Finally, by \cite{Mer_Weiss03}, the trivial scan is optimal for scandiction of noise-free Gaussian random fields. By Corollary \ref{coro. noisy scand. gaussian} herein, this is also the case in scandiction of Gaussian fields corrupted by Gaussian noise. Whether the same hold for scanning and \emph{filtering} of Gaussian random fields corrupted by Gaussian noise remains unanswered.

\appendix\section{Appendixes}
%
\subsection{Proof of Theorem \ref{theo. bound on sens. of scantering Arbitrary}}\label{app. proof of bound on sens. of scantering Arbitrary}
The proof resembles the proof of Theorem \ref{theo. bound on sens. of scantering Arbitrary}. However, the derivations leading to the analogue of \eqref{eq. proof of scantering bound G} are slightly different. For any input field $X_{V_n}$, we have
\begin{align} 
\frac{1}{n^2}E_{Q_{V_n}}&L_{(\Psi,\tilde{F}^{opt})}\left(X_{V_n},Y_{V_n}\right)
\nonumber\\ 
&= \frac{1}{n^2}\sum_{i=1}^{n^2}{\Var\left(X_{\Psi_i}|Y_{\Psi_1}^{\Psi_{i}}\right)}
\nonumber\\
&= \frac{1}{n^2}\sum_{i=1}^{n^2}\Bigg\{\int_0^1{\Var\left(X_{\Psi_i}|Y_{\Psi_1}^{\Psi_{i-1}},\{Y_t^{(c)}\}_{t \in [i-1,i-1+t]}\right)\der t}  
\nonumber\\
& \qquad - \Bigg[ \int_0^1{\Var\left(X_{\Psi_i}|Y_{\Psi_1}^{\Psi_{i-1}},\{Y_t^{(c)}\}_{t \in [i-1,i-1+t]}\right)\der t} - \Var\left(X_{\Psi_i}|Y_{\Psi_1}^{\Psi_{i}}\right)\Bigg] \Bigg\}
\nonumber\\
&\stackrel{(a)}{\ge} \frac{1}{n^2}\sum_{i=1}^{n^2}\int_0^1{\Var\left(X_{\Psi_i}|Y_{\Psi_1}^{\Psi_{i-1}},\{Y_t^{(c)}\}_{t \in [i-1,i-1+t]}\right)\der t}  
 -f^*\left(X_{V_n},\sn\right)
\nonumber\\
&= \frac{1}{n^2}2\sn I \left(X_{V_n};Y_{V_n}\right) - f^*\left(X_{V_n},\sn\right),
\end{align}
where (a) results from the definition of $f^*()$. The rest of the proof follows similar to the proof of Theorem \ref{theo. bound on sens. of scantering Arbitrary}, since for any $X_{V_n}$ and $\sn$ it is clear that $f^*(X_{V_n},\sn)$ is non negative.
\subsection{Proof of Lemma \ref{lem. scandictability for additive and
    square error}}\label{app. proof of lemma scandictability for
  additive and square error}Without loss of generality, we assume $EN_t=0$. From Proposition \ref{prop. existance of the limit}, we have,
\begin{equation}
\bar{U}(\rho_s,Q)=\lim_{n \rightarrow \infty}\inf_{(\Psi,F) \in \calS(B_n)}
E_{Q_{B_n}}\frac{1}{|B_n|}\sum_{t=1}^{|B_n|}
{\left(X_{\Psi_t}-F_t(y_{\Psi_1},\ldots,y_{\Psi_{t-1}})\right)^2}.
\end{equation}
However, since $Y_{\Psi_t} = X_{\Psi_t} + N_{\Psi_t}$, and $N_{\Psi_t}$ is
independent of $N_{\Psi_{t'}}$, $t' \neq t$ and of all $\{X_t\}$, we have
\begin{eqnarray}
E_{Q_{B_n}}\frac{1}{|B_n|}\sum_{t=1}^{|B_n|}
{\left(X_{\Psi_t}-F_t(y_{\Psi_1},\ldots,y_{\Psi_{t-1}})\right)^2} &&
\nonumber\\
&& \hspace{-5cm} = E_{Q_{B_n}}\frac{1}{|B_n|}\sum_{t=1}^{|B_n|}
{\left(Y_{\Psi_t}-F_t(y_{\Psi_1},\ldots,y_{\Psi_{t-1}})-N_{\Psi_t}\right)^2}
\nonumber\\
&& \hspace{-5cm} = E_{Q_{B_n}}\frac{1}{|B_n|}\sum_{t=1}^{|B_n|}
\bigg\{\left(Y_{\Psi_t}-F_t(y_{\Psi_1},\ldots,y_{\Psi_{t-1}})\right)^2
\nonumber\\
&& \hspace{-5cm} \qquad -2N_{\Psi_t}\left(X_{\Psi_t}+N_{\Psi_t}-F_t(y_{\Psi_1},\ldots,y_{\Psi_{t-1}})\right)+N_{\Psi_t}^2\bigg\}
\nonumber\\
&& \hspace{-5cm} = E_{Q_{B_n}}\frac{1}{|B_n|}\sum_{t=1}^{|B_n|}
\left(Y_{\Psi_t}-F_t(y_{\Psi_1},\ldots,y_{\Psi_{t-1}})\right)^2 - \sn.
\end{eqnarray}
That is,
\begin{equation}
U(\rho_s,Q)=\lim_{n \rightarrow \infty}\inf_{(\Psi,F) \in \calS(B_n)}
E_{Q_{B_n}}\frac{1}{|B_n|}\sum_{t=1}^{|B_n|}
\left(Y_{\Psi_t}-F_t(y_{\Psi_1},\ldots,y_{\Psi_{t-1}})\right)^2 - \sn,
\end{equation}
which completes the proof.
%
\subsection{The Martingale Property of $\left(\Delta_{(\Psi,F)}(x_B,y_B)_t,\calF^\Psi_t\right)$}\label{app. proof of martingale}
The proof follows that of \cite[Lemma 1]{Weiss_Mer01}. However, notice that due to the data-dependent scanning $\calF^\Psi_t$ is not generated by a \emph{fixed} set of random variables, that is, over a fixed set of sites, but by a set of $t$ random variables which may be different for each instantiation of the random field (as for each $t$, $\Psi_t$ depends on $Y_{\Psi_1}^{\Psi_{t-1}}$). Yet, the expectation will always be with respect to the random variables \emph{seen so far}.

By \eqref{def. Delta},
\begin{equation}
\Delta_{(\Psi,F)}(x_B,y_B)_t = \sum_{i=1}^{t}\left(h(y_{\Psi_i})-x_{\Psi_i}\right)l_0(F_i(y_{\Psi_1}^{\Psi_{i-1}}))+\sum_{i=1}^{t}\left(x_{\Psi_i}-h(y_{\Psi_i})\right)l_1(F_i(y_{\Psi_1}^{\Psi_{i-1}})).
\end{equation}
Defining
\begin{equation}
m_t \defined \sum_{i=1}^{t}\left(h(Y_{\Psi_i})-x_{\Psi_i}\right)l_0(F_i(Y_{\Psi_1}^{\Psi_{i-1}})),
\end{equation}
we have,
\begin{eqnarray}
E\left\{m_{t+1}|\calF^\Psi_t\right\} &=& E\left\{\sum_{i=1}^{t+1}\left(h(Y_{\Psi_i})-x_{\Psi_i}\right)l_0(F_i(Y_{\Psi_1}^{\Psi_{i-1}}))|\calF^\Psi_t\right\}
\nonumber\\
&=& E\left\{\left(h(Y_{\Psi_{t+1}})-x_{\Psi_{t+1}}\right)l_0(F_{t+1}(Y_{\Psi_1}^{\Psi_t}))|\calF^\Psi_t\right\}
\nonumber\\
&&\hspace{1cm}+ E\left\{\sum_{i=1}^{t}\left(h(Y_{\Psi_i})-x_{\Psi_i}\right)l_0(F_i(Y_{\Psi_1}^{\Psi_{i-1}}))|\calF^\Psi_t\right\}
\nonumber\\
&=& E\left\{\left(h(Y_{\Psi_{t+1}})-x_{\Psi_{t+1}}\right)|\calF^\Psi_t\right\} l_0(F_{t+1}(Y_{\Psi_1}^{\Psi_t}))
\nonumber\\
&&\hspace{1cm}+ \sum_{i=1}^{t}\left(h(Y_{\Psi_i})-x_{\Psi_i}\right)l_0(F_i(Y_{\Psi_1}^{\Psi_{i-1}}))
\nonumber\\
&=& E\left(h(Y_{\Psi_{t+1}})-x_{\Psi_{t+1}}\right) l_0(F_{t+1}(Y_{\Psi_1}^{\Psi_t}))+ m_t
\nonumber\\
&=& m_t,
\end{eqnarray}
where the third equality is since $Y_{\Psi_{t_0}}$ is $\calF^{\Psi}_t$ measurable for any $t_0 \leq t$, the fourth is since $h(Y_{\Psi_{t+1}})-x_{\Psi_{t+1}}$ is independent of $\calF^{\Psi}_t$ and the fifth is since $h(Y_{\Psi_{t+1}})$ is an unbiased estimate for $x_{\Psi_{t+1}}$. Hence, $(m_t,\calF^\Psi_t)$ is a zero-mean martingale (note that $Em_1=0$). Analogously, $\sum_{i=1}^{t}\left(x_{\Psi_i}-h(y_{\Psi_i})\right)l_1(F_i(y_{\Psi_1}^{\Psi_{i-1}}))$ is also a zero-mean martingale with respect to $\calF^\Psi_t$, which completes the proof.

\bibliographystyle{../latex/IEEEbib}
\bibliography{../latex/prediction_and_coding}
\end{document}